\documentclass[aip,floatfix,amsmath,amssymb,reprint]{revtex4-1}
\usepackage{graphicx}
\usepackage{dcolumn}
\usepackage{bm}

\usepackage[utf8]{inputenc}
\usepackage[T1]{fontenc}
\usepackage{mathptmx}
\usepackage{etoolbox}
\usepackage{subfigure}
\usepackage{comment}
\usepackage{orcidlink}
\usepackage{placeins}

\begin{document}


\title{Instability of a Counter-Streaming Collisionless Pair Plasma I: Evolution and Parameterization of the Weibel Instability} 



\author{Michael C. Sitarz \orcidlink{0000-0001-9003-0737}}
\affiliation{Department of Physics and Astronomy, University of Kansas, Lawrence, KS 66045}
\email{mcsitarz@ku.edu}


\date{April 23, 2024}

\begin{abstract}
Energetic astrophysical phenomena, such as $\gamma$-ray bursts and supernova explosion-driven shocks in collisionless plasmas, involve various plasma kinetic instabilities, such as the Weibel instability. In this paper, we explore the evolution of the Weibel instability using the spectral analysis and physical field evolution of particle-in-cell simulations. Effects of varying parameters of the initial counter-streaming beams are detailed in reference to the total evolutionary timeline of the instability. Generation of a secondary instability and excitation of electromagnetic waves primary decay have been observed. We also discuss the filament mergers during the instability decay.
\end{abstract}

\maketitle 

\section{Introduction}
\indent High energy events are commonplace within the universe. These events exist in different plasma environments like the interplanetary medium (IPM), the interstellar medium (ISM), the intergalactic medium (IGM), or the intercluster medium (ICM). The phenomena themselves come in a variety of sizes, energies, and locations: supernova blast waves (SN) \cite{1}, IPM shocks \cite{2}, quasar jets \cite{2,3}, solar flares \cite{4}, pulsar wind nebula (PWN) \cite{5}, relativistic jets in active galactic nuclei (AGN), gamma ray bursts (GRB) \cite{6}, and the virilization of IGM. The commonalities these events share is the formation of collisionless shock environments and the production of plasma and associated turbulence. 

\indent Weibel \cite{7} and Fried \cite{8} demonstrated that magnetic fields can be generated in a current-free collisionless plasma via an instability driven by particle distribution function (PDF) anisotropy. This can be brought forth by a temperature or a velocity gradient, generating a runaway magnetic field. One of the most common set-ups of this system is the two counter streaming beam Weibel instability. Since then, much work has been done using the Weibel instability (for example, Jitter radiation produced as a by-product of particle acceleration during filament building \cite{9}). While there is a plethora of knowledge about the instability ignition, general environment, and electromagnetic mechanisms within the instability, there is still question about how the Weibel instability dissipates and evolves after the saturation point of the filaments. This paper puts forth an answer this question. 

\indent The Weibel instability is excited by an anisotropy in either temperature or velocity in an otherwise uniform environment. The anisotropy in the system generates a perturbation in the form which initializes the instability and run-away magnetic field. The lifetime of the Weibel instability can be characterized by distinct eras or epochs along its evolution. The final stages of this evolution can be described by the saturation of the filamentary structure of the system, and the subsequent coalescence of these filaments. Generally, violent mergers occur, accelerating particles and generating energy and radiation. We show that during violent filament mergers, fields are exited and the dissipation of the instability system is interrupted. 

\indent This paper presents a physical and fundamental connection between the system's physical parameters and excitation of physical fields. We also present a discussion on the spectral waves generated during this process. This connection can be shown as a result of first principle simulations, field analysis in the space-time Fourier domain, and $k-\omega$ dispersion plot studies. The Weibel instability generated by two counter streaming beams in time from the ``cold beam”  system evolves into a ``warm plasma” distribution, where the momenta can be described by Maxwell-J{\"u}ttner Distribution \cite{10}. It is at this critical point of saturation of the filaments where the Weibel instability approaches breakdown. As the plasma transforms into the ``warm” regime through the dissipation and merger of the saturated filaments that the fields are excited by locally accelerated particles. 

\indent The remainder of the paper will be organized as follows: \S$2$ is a review of the Weibel instability, \S$3$ is a description of the simulation set-up and a discussion of the analytical techniques used in the study of the data, and \S$4$ will discuss the results of the analysis and the conclusions drawn from the study. Finally, \S$5$ will contain concluding remarks and possible implications.

\section{Major Instability Background}
\subsection{Weibel Instability Timeline}
\indent The Weibel instability (WI) is proposed to be the source of intense magnetic fields within the GRB prompt emission, afterglow \cite{9} and astrophysical shock frames \cite{11}. These claims were later proved numerically by a number of studies \cite{12,13,14,15,16}. The magnetic field turbulence generated by the WI operating at the shock front \cite{9} is of a few ion skin depths \cite{17}, is sub-Larmor in scale \cite{18}, and is an important mechanism in the mediation of the GRB shock \cite{19}. This small scale turbulence is not uncommon in an astrophysical setting, with other examples being the electromagnetic Whistler, filamentation, and mixed modes or electrostatic (ES) Langmuir oscillations \cite{9}. 

\indent The WI was derived by Weibel in $1959$ using a fully kinetic analysis \cite{7}, where he considered a non-relativistic plasma with an anisotropic particle distribution function (PDF). It begins as a very weakly or non-magnetized plasma with an anisotropic velocity distribution of electrons and fixed ions, with the electron temperature dependency based on direction (bi-Maxwellian anisotropy). 
\begin{equation}\label{Eq:Bimax_Dist}
    f(v) = \frac{n}{\sqrt{\pi^3}\Theta^2_\perp\Theta^2_\parallel}exp\left[-\left(\frac{v^2_\perp}{\Theta^2_\perp} + \frac{v^2_\parallel}{\Theta_\perp^2}\right)\right],
\end{equation}
where the $\Theta$ terms are defined as (noting that $T_\perp \ne T_\parallel$).
\begin{equation}\label{Eq:Theta_Temp}
    \Theta_{\parallel / \perp} = \sqrt{\frac{2kT_{\parallel / \perp}}{m}}.
\end{equation}
The total distribution for a particle species $s$ can be given as the sum of equilibrium and perturbed PDFs 
\begin{equation}\label{Eq:Total_Dis_Species}
    f_s = f_{0s} + \hat{f}_s.
\end{equation}
This then generates a transverse mode of perturbations in the electrons.

\begin{figure}[h]
    \centering
    \includegraphics[scale=0.08]{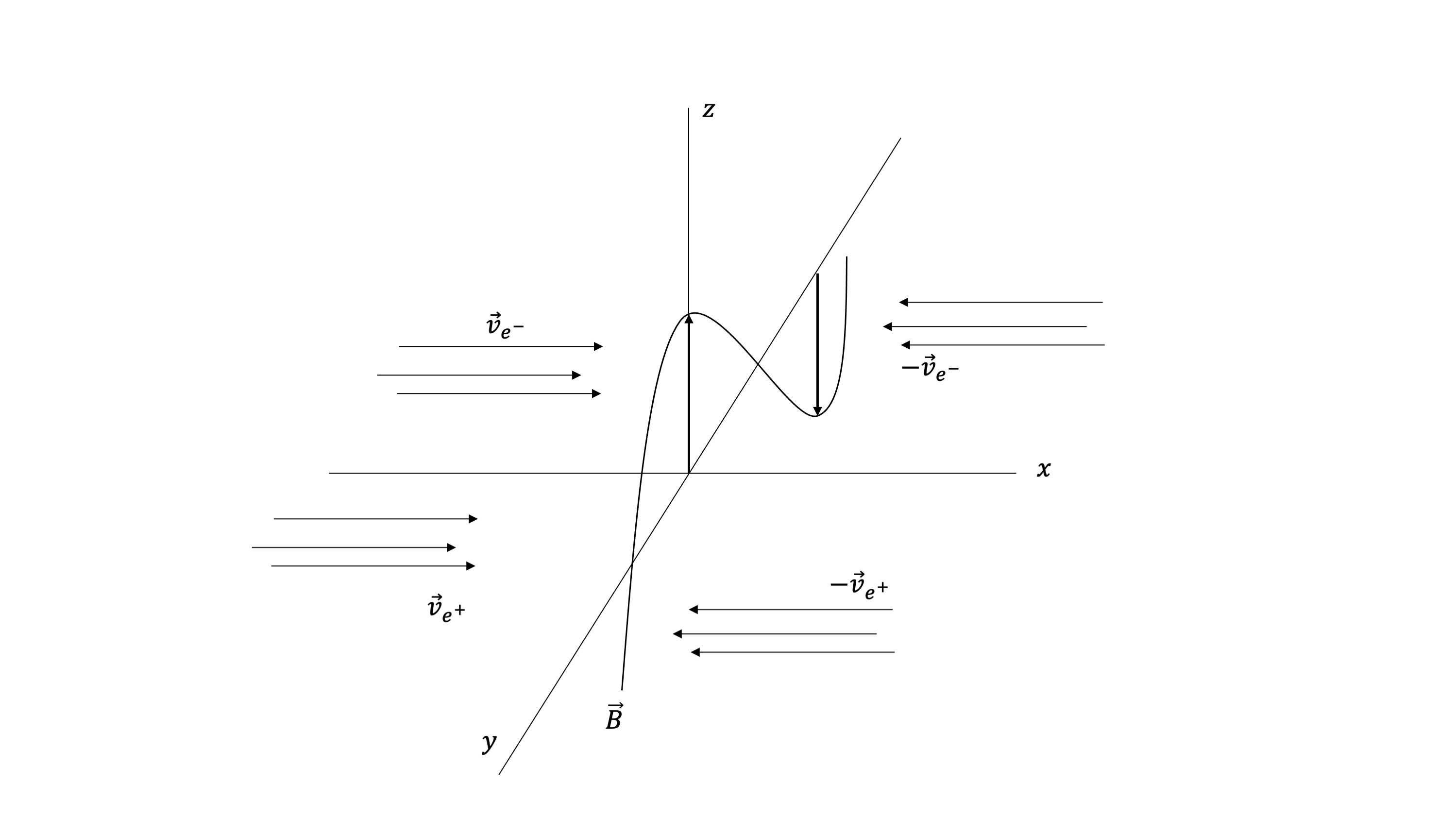}
    \caption{Example set-up of two counter streaming beams with a sinusoidal mode $\Vec{B}$ field normal to the streaming plane. The beams are composed of both positrons and electrons with $n_{e^-} = n_{e^+}$.}
    \label{Fig:Weibel_Cartoon}
\end{figure}

\begin{figure}[h]
    \centering
    \includegraphics[scale=0.08]{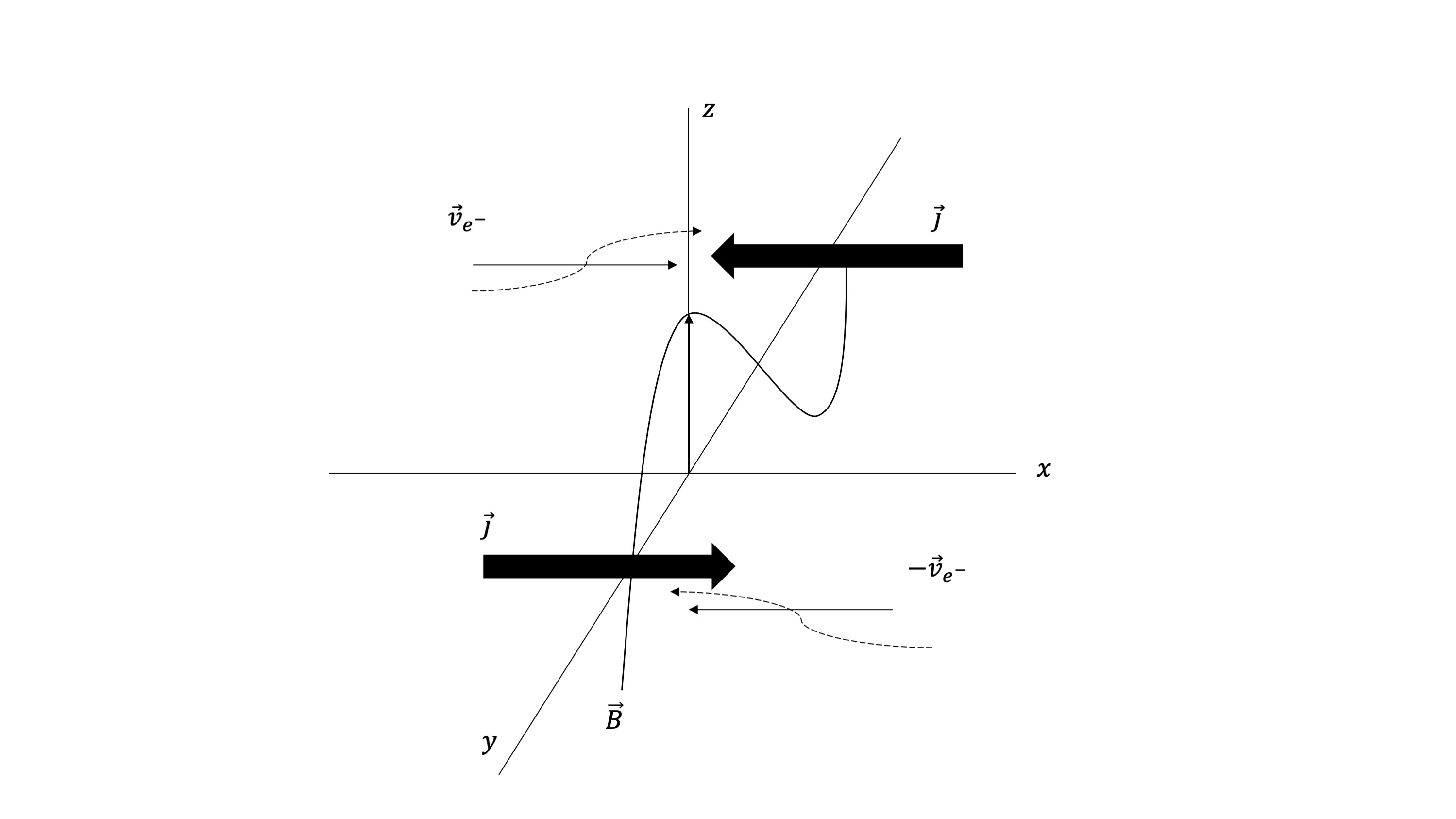}
    \caption{As the beams stream into the $\Vec{B}$ field, the particles' trajectory begins to deflect towards the nodes of the field. These particles bunches then produce the signature current filaments in the direction of flow.}
    \label{Fig:Filament_Cartoon}
\end{figure}

\indent Fried \cite{8} treated an anisotropic PDF more specifically as a two stream in a cold plasma. He describes counter streaming planes of electrons with a very small sinusoidal magnetic field normal to the streaming plane, producing a runaway effect (Fig. \ref{Fig:Weibel_Cartoon}) This sinusoidal field is now treated as initial fluctuations in the system.
The Lorentz force, $\vec{F} = \frac{e}{c}(\vec{v} \times \vec{B})$, acts on the charged particles and deflects their trajectories. This results in spatial concentrations of particles that generate separate current filaments. $\vec{B}$ in the filaments increases the initial fluctuation (Fig. \ref{Fig:Filament_Cartoon}) 
The maximum growth rate of the instability
\begin{equation}\label{Eq:Max_Growth_Rate}
    \Gamma_{max} \approx \gamma^{\frac{1}{2}}\omega_p \rightarrow \frac{c}{\lambda_{De}},
\end{equation}
and the fastest growing mode 
\begin{equation}\label{Eq:Fastest_Mode}
    k_B = \frac{\omega_{p,s}}{c}
\end{equation}
with plasma frequency $\omega_p = \sqrt{\frac{4\pi e^2}{m}}$,
and Debye length $\lambda_{De} = \frac{v_{th}}{\omega_p} = \sqrt{\frac{kT_e}{4\pi ne^2}}$ set the correlation scale of the produced magnetic fields, and Lorentz factor $\gamma = \frac{1}{\sqrt{1 - (v/c)^2}}$. 

\begin{figure}[h]
    \centering
    \includegraphics[scale=0.4]{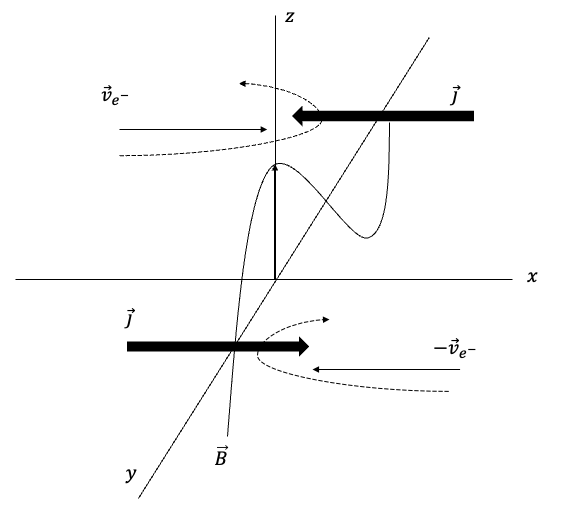}
    \caption{Cartoon depicting the saturation point of the WI. When the bunches of particles at the nodes of the field reach a maximum point, the current filaments cease to grow as particles are completely deflected away from the nodes.}
    \label{Fig:Filament_Schematic}
\end{figure}

\begin{figure}[ht]
    \centering
    \includegraphics[scale=0.5]{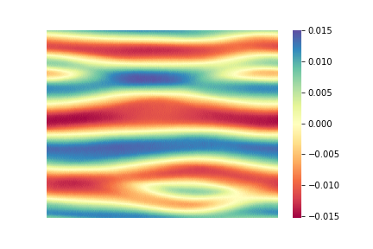}
    \caption{Field heat map depicting the magnetic fields ($B_z$) at saturation point just before they begin to dissipate. These filaments (blue)  will no longer grow in size, but begin to slowly drift towards each other until merging.}
    \label{Fig:Filament_Fields}
\end{figure}

\indent The Lorentz force continues to deflect particles orbits in larger quantities, which then amplify the magnetic field in a cyclical nature. These deflections are on the scale of the Larmor radius $\rho_L = v_{\perp,B}/\omega_{c,s}$ (with cyclotron frequency $\omega_{c,s} = \frac{eB}{m_sc}$). As the magnetic field increases, so do the deflection paths. This suppresses their free streaming across field lines. When the field reaches $k_b\rho_L \sim 1$, particles become magnetically trapped and field amplification ceases. The system has now reached saturation and begins to breakdown (Figs. \ref{Fig:Filament_Schematic} and \ref{Fig:Filament_Fields})

\subsection{Current Filament Dynamics}
\indent It is important to discuss the dynamics and evolution of the current filaments present in the WI system, as they are the most salient features of the instability process. This discussion is derived from \cite{20} and the sources therein. At the beginning of the instability process, as particles begin to deflect and bunch, small filaments form within the box. This example will consider one-dimensional filaments, but the theory is identical in higher dimensions without loss of generality. We assume that all filaments start out identical in size and quantity (charge and current neutrality must be preserved). They possess initial characteristics as follows: diameter $D_0$, mass per unit length $\mu_0 \simeq 0.25mnD_0^2\pi$ (particle mass $m$ and number density $n$), current $I_0$, and spatial separation $d_0 \simeq 2D_0$ from center to center. They form randomly distributed throughout the box and at rest but vulnerable to attraction/repulsion forces.

\indent Filaments grow as more and more particles are deflected into B-field nodes and amplify the field. The filaments will grow in transverse scale beginning with the skin length scale, $\lambda_B \approx 2\pi c/\omega_{p,e^-}$, and continues to scale with the correlation length of the produced fields \cite{20}. As plasma temperature increases, the growth rate decreases \cite{21} but is not limited and can grow unconstrained. The linear theory gives no good length prediction for the transverse filament scale (parallel to beam direction). At saturation, like-current filaments begin to drift towards each other, orienting into a more regular pattern. This is accompanied by a decrease in the magnetic field associated with the topological change in the box. 

\indent The spatial distribution of filaments appears as a regular pattern post saturation, but they are not stationary. The equation of motion for the filaments can be easily found. The magnetic fields produced by a filament is given by 
\begin{equation}\label{Eq:Filament_Mag_Field}
    B_0(r) = \frac{2I_0}{cr},
\end{equation}
with cylindrical radius $r$. Assuming the filament is straight (no kinks or bends), the force per unit length is given as
\begin{equation}\label{Eq:Filament_Force_Length}
    \frac{dF}{dl} = \frac{-B_0I_0}{c}.
\end{equation}
Since,
\begin{equation}\label{Eq:Fil_Force_Length_Reduced}
    \frac{dF}{dl} = \mu \Ddot{x},
\end{equation}
where reduced mass $\mu = \mu_0/2$, we have:
\begin{equation}\label{Eq:Filament_EoM}
    \Ddot{x} = \frac{-2I_0^2}{c^2\mu x},
\end{equation}
with $r = 2x$.

\indent With the filament movement modeled, the merger time can now be estimated. Filament merger is first defined as when two filaments touch, as opposed to a full integration. This occurs at $d_0 \simeq D_0$. The time scale of filament mergers is independent of the merger process itself, as the coalescence of two current filaments involves complex current redistribution dynamics. The force interaction between filaments gets weaker as distance increases, limiting merger rate. The time scale $\tau$ can be readily estimated from $\Ddot{x}$ given that $\Ddot{x} \sim (d_0/2)/\tau_0^2$. For non-relativistic motion
\begin{equation}\label{Eq:NR_Fil_Merge_Time}
    \tau_{0,NR} \sim \sqrt{\frac{D_0^2c^2\mu_0}{2I_0^2}},
\end{equation}
where the max velocity of a filament is limited to 
\begin{equation}\label{Eq:Fil_Max_Velo}
    \vec{v}_{Max,0} \sim \frac{D_0}{2\tau_0} \sim \frac{I_0}{\sqrt{2c^2\mu_0}}.
\end{equation}
If filament motion becomes comparable to $c$, i.e. $t(x) \simeq \frac{x}{c}$, then the regime becomes relativistic and the time scale now reads
\begin{equation}\label{Eq:R_Fil_Merge_Time}
    \tau_{0,R} \simeq \frac{d_0c}{2} = \frac{D_0}{c}.
\end{equation}

\indent Filament merging is a hierarchical and self similar process. If the system initially has $N_0$ number of filaments, filaments will merge pairwise under the given time scales into a new ($1^{st}$) generation of filaments numbering $N_0/2$. The properties of the initial filaments scale as the $k^{th}$ generation is produced, following
\begin{multline}\label{Eq:Number_of_kth_Gen_Fil}
    I_k = 2^kI_0, \ \mu_k = 2^k\mu_0, \ D_k = 2^{k/2}D_0, \\
    d_k \sim \frac{D_k}{2}, \ \tau_{k,NR} = \tau_{0,NR}, \\
    \tau_{k,R} = 2^{k/2}\tau_{0,R}.
\end{multline}

\indent Filament motion is not constrained in the non-relativistic or relativistic regimes. Velocities will increase with each generation merged, following $\vec{v}_{Max,k} = 2^{k,0}\vec{v}_{Max,0}$. Non-relativistic can become relativistic following  $j = 2log_2(c/\vec{v}_{Max,0})$ mergers. 

\indent Finally, filament parameters and correlation scales can be expressed as a function of physical time. Based on merger levels, it will take $t = \sum^k_{k'=0} \tau_{k'}$ to complete the $k$ mergers. Using this relation, merger level can be shown as
\begin{equation}\label{Eq:Fil_Merge_Level}
    k \simeq 2log_2(\frac{t}{\tau{0,R}}),
\end{equation}
with characteristic magnetic field length
\begin{equation}\label{Eq:Char_Fil_Mag_Length}
    \lambda_B(t) = D_02^{t/\tau_{0,NR}}.
\end{equation}

\subsection{Analytical Dispersion Relation}
\indent The following derivation is a mix of sources including \cite{10, 22,23,24}, we encourage the reader to consult their papers for a more detailed viewing.  While the system in this study is a $e^\pm$ pair background plasma with $e^\pm$ beams, a more general dispersion relation derivation follows using electrons and ions as the plasma composition. This derivation is meant to bring the reader up to speed with a more general example before analyzing a more specific system. Full derivation using the system analyzed in this study will be published in follow-up publications. 

\indent The initial set-up contains a uniform plasma with immobile and neutralizing ion background and two counter streaming $e^-$ beams. The ion background has density $n_i = \sum_\alpha n_{o,\alpha}$ over $\alpha$ species of ion. For this derivation, ``ion'' is simply a particle that is not an electron. The beams will propagate in a $2D$ box along the $\hat{x}$ direction with unperturbed number density $n_{0,\alpha}$, with $\alpha$ denoting the electron species (beam population). Together with their velocities $v_{0,x,\alpha}$, the beams contribute zero global current density
\begin{equation}\label{Eq:Beam_Current_Density}
    \sum_\alpha n_{0,\alpha} v_{beam,x,\alpha} = 0.
\end{equation}
The three-momenta of the beams can be further defined by
\begin{equation}\label{Eq:Beam_Three_Momenta}
    \vec{v}_\alpha = \frac{\vec{p}_\alpha c}{\sqrt{m^2c^2 + p_\alpha^2}}.
\end{equation}

\indent Begin with Maxwell's equations
\begin{equation}\label{Eq:Gauss_Law_Max}
    \nabla \cdot \vec{E} = \sum_\alpha n_{beam,\alpha},
\end{equation}
\begin{equation}\label{Eq:No_Mono_Max}
    \nabla \cdot \vec{B} = 0,
\end{equation}
\begin{equation}\label{Eq:Faraday_Law_Max}
    \nabla \times \vec{E} = \frac{-\partial B}{\partial t},
\end{equation}
and
\begin{equation}\label{Eq:Ampere_Law_Max}
    \nabla \times \vec{B} = \sum_\alpha n_{beam,\alpha} \vec{v}_{beam,x,\alpha} + \frac{1}{c^2}\frac{\partial \vec{E}}{\partial t},
\end{equation}
the relativistic dynamics of $e^-$
\begin{equation}\label{Eq:Rel_Momenta_Electron_MHD}
    \frac{\partial \vec{p}_\alpha}{\partial t} + (\vec{v}_\alpha \cdot \nabla)\vec{p}_\alpha = -e\left(\vec{E} + \frac{\vec{v}_\alpha}{c} \times \vec{B}\right),
\end{equation}
\begin{equation}\label{Eq:Continuity_Electron_MHD}
    \frac{\partial n_{beam,\alpha}}{\partial t} + \nabla \cdot n_{0,\alpha}\vec{v}_\alpha = 0,
\end{equation}
and vector potential
\begin{equation}\label{Eq:Mag_Vector_Potential}
    \vec{B} = \nabla \times \vec{A},
\end{equation}
\begin{equation}\label{Eq:Momenta_due_to_Vector_Potential}
    p_{l,\alpha} - \frac{eA_l}{c} = p_{beam,l,\alpha},
\end{equation}
where $l$ represents a coordinate ($x$, $y$, $z$). The following substitution has been made for explicitness 
\begin{equation}\label{Eq:Current_Density_Beam}
    \vec{j} = -n_{beam,\alpha}\vec{v}_{beam,x,\alpha}.
\end{equation}

\indent Coupling the $e^-$ momentum and density with the Maxwell equations recovers the following without loss of generality with coordinates ($l$, $m$, $n$)
\begin{equation}\label{Eq:Max_Gauss_Part}
    \frac{\partial E_l}{\partial l} = 4\pi e\left(n_i - \sum_\alpha n_\alpha\right),
\end{equation}
\begin{equation}\label{Eq:Max_Ampere_Part}
    \frac{\partial B_n}{\partial m} = -\frac{4 \pi e}{c}\sum_\alpha n_\alpha v_{x,\alpha} + \frac{1}{c}\frac{\partial E_l}{\partial t},
\end{equation}
\begin{equation}\label{Eq:Max_Double_Faraday_Part}
    -\frac{\partial^2 E_l}{\partial m^2} = -\frac{1}{c}\frac{\partial}{\partial t}\frac{\partial B_n}{\partial m},
\end{equation}
\begin{equation}\label{Eq:Max_Faraday_Part}
    \frac{1}{c}\frac{\partial B_n}{\partial t} = \frac{\partial E_l}{\partial m},
\end{equation}
\begin{equation}\label{Eq:Momenta_Part}
    \frac{\partial p_{l,\alpha}}{\partial t} + v_{m,\alpha}\frac{\partial p_{m,\alpha}}{\partial m} = -e \left(E_m - \frac{v_{x,\alpha}}{c}B_n \right),
\end{equation}
\begin{equation}\label{Eq:Continuity_Part}
    \frac{\partial n_\alpha}{\partial t} + \frac{\partial n_\alpha v_{m,\alpha}n_{m,\alpha}}{\partial m} = 0.
\end{equation}

\indent The above equations are now linearized using a small plane wave perturbation of the form 
\begin{equation}\label{Eq:Linear_Plane_Wave}
    F(x,y,t) = fexp[ik_xx + ik_yy - i\omega t],
\end{equation}
applied to the velocities, densities, and fields. $\omega$ represents the angular frequency perturbation while $k_i$ represents the wave-vector perturbation. In the linearization calculations, the following relationships are used: $\frac{\partial}{\partial t} \rightarrow -i\omega$ and $\nabla \rightarrow ik_y + ik_x$.

\indent For the remainder of this example, we will only be concerned with variables with respect to $y$ and $t$. This truncation of dimensions is valid due to the nature of the WI and the Two Stream instability (TSI). When $k_y = 0$, the TSI arises from a plane wave perturbation. If there is oblique and intermediate propagation angles where $k_x$ and $k_y$ are non-vanishing, the WI and the TSI are coupled into a single branch. In one dimension, the TSI has a set cutoff at $k_x^{Max}$ beyond which the TSI is no longer unstable. The full, $3$-space dispersion can be found in the mentioned sources and will be examined in depth in subsequent publications.

\indent Linearizing (Eqs. \ref{Eq:Max_Gauss_Part}, \ref{Eq:Max_Ampere_Part}, \ref{Eq:Max_Double_Faraday_Part}, \ref{Eq:Max_Faraday_Part}, \ref{Eq:Momenta_Part}, \ref{Eq:Continuity_Part}) and substituting them into Eq. (\ref{Eq:Max_Faraday_Part}) recovers a sixth order  relation for the WI
\begin{equation}\label{Eq:6th_Order_DR}
    (\omega^2 - \Omega_a^{2})[\omega^4 - \omega^2(k^2c^2 + \Omega_b^2) - k^2c^2\Omega_c^2] - k^2c^2\Omega_d^2 = 0.
\end{equation}
Here, the following substitutions are used 
\begin{equation}\label{Eq:DR_Omega_a}
    \Omega_a^2 = \omega_{pe}^2 \sum_\alpha \frac{n_{beam,\alpha}}{n_i \Gamma_\alpha},
\end{equation}
\begin{equation}\label{Eq:DR_Omega_b}
    \Omega_b^2 = \omega_{pe}^2 \sum_\alpha \frac{n_{beam,\alpha}}{n_i \Gamma_\alpha^3},
\end{equation}
\begin{equation}\label{Eq:DR_Omega_c}
    \Omega_c^2 = \omega_{pe}^2 \sum_\alpha \frac{n_{beam,\alpha} v_{beam,x,\alpha}^2}{n_i \Gamma_\alpha c^2},
\end{equation}
\begin{equation}\label{Eq:DR_Omega_d}
    \Omega_d^2 = \omega_{pe}^2 \sum_\alpha \frac{n_{beam,\alpha} v_{beam,x,\alpha}}{n_i \Gamma_\alpha c},
\end{equation}
where
\begin{equation}\label{Eq:Lorentz_Factor_Beam}
    \Gamma_\alpha = \left(1 - \frac{v_{beam,x,\alpha}^2}{c^2}\right)^{-1/2}.
\end{equation}
For a $e^\pm$ beam, we can further denote the beam density as 
\begin{equation}\label{Eq:Beam_Density_DR}
    n_{beam, e^-} = n_{beam, e^+} = n_{beam}/2,
\end{equation}
with velocities
\begin{equation}\label{Eq:Beam_Velocity_DR}
    v_{beam, e^-} = v_{beam, e^+} = v_{beam,x}/2.
\end{equation}
We may also remove the ion background $n_i$ without disrupting the following relations, as it was immobile and present for charge neutrality. 

\indent This dispersion relation can be solved by using substitution ($u = \omega^2$), which reduces the order of function, and then employing the companion matrix method to find the eigenvalues (roots). Three branches can be found from the eigenvalues of the function. Two real ($\mathbb{C} = 0$), oscillatory modes and a single exponentially growing mode with imaginary components ($\mathbb{C} \ne 0$). This exponentially growing mode is the WI. The maximum growth rate $\Gamma_{Max}$ is found in the short wave limit where $k^2c^2$ dominates over the $\Omega_i$ terms. This $\Gamma$ is not the same as $\Gamma_\alpha$, which is used for kinematics. When the growth rate is discussed, the $\Gamma$ will be explicitly mentioned to avoid confusion. Applying this condition to Eq. (\ref{Eq:6th_Order_DR}) finds
\begin{equation}\label{Eq:WI_Growth_DR}
    \Gamma_{Max} \approx \frac{\sqrt{\sqrt{(\Omega_a^2 + \Omega_c^2)^2 - 4\Omega_d^4} - (\Omega_a^2 - \Omega^2_c)}}{\sqrt{2}}.
\end{equation}
Long wave dependence ($k^2c^2 \sim 0$) gives
\begin{equation}\label{Eq:WI_Growth_DR_LongWave}
    \Gamma_{Max} \approx \sqrt{\frac{\Omega_a^2\Omega_c^2 - \Omega_d^4}{\Omega_a^2\Omega_b^2}}.
\end{equation}

\section{Analysis of Simulation Data}
\subsection{Simulation Set-Up}
\indent The analysis is done by running state of the art PIC simulations of two counter streaming electron-positron beams using the code (TRISTAN-MP \cite{TMP}). Four separate parametrizations were simulated. The fiducial simulation had a beam propagation Lorentz factor of $3$, a particle density of $64$ particles per cell, and an electron skin depth of $32$ cells. This simulation is labeled $S1$. To test the convergence of simulation results and study the effect of different parameters on the system, three additional simulations were performed as follows in reference to the fiducial simulation: halving the Lorentz factor of the beams ($S2$), doubling the particle per cell density ($S3$), or dividing the skin depth by $8$ ($S4$). A full table of simulation parameters can be found in Appendix C.

\indent The field data ($B_z$, $E_x$, and $E_y$) from each simulation is then split into $15$ epochs, with each epoch containing $256$ snap shots. This snap shot number per epoch is a requirement of the fast Fourier transform (FFT) spectral analysis and the mirror expansion performed. To ensure the signal is periodic, the mirror of the field is taken along the time axis, creating an array ($2 \times t, x, y)$ that is transformed into ($\omega, k_x, k_y$) with the signal in the omega repeating after $\omega_{max}/2$. The physical data either then plotted directly and studied using energy evolution or transformed into spectral data. The spectral data is used to plot the full dispersion relation and the dispersion relation expansion. The full dispersion relation is a plot of $k$ vs. $\omega$ for the $Log(Amplitude)$ of the FFT field analyzed. The expansion is a $k_x$ vs. $k_y$ plot for each $\omega$ value in the system.

\subsection{Magnetic Energy Evolution}
\begin{figure*}
    \centering
    \begin{subfigure}
        \centering
        \includegraphics[scale=0.4]{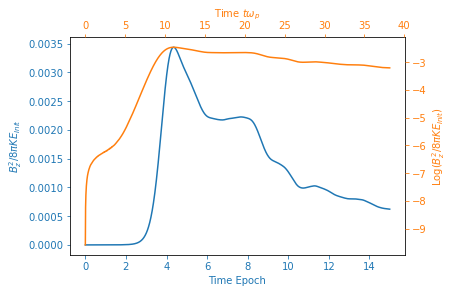}
    \end{subfigure}
    \begin{subfigure}
        \centering 
        \includegraphics[scale=0.4]{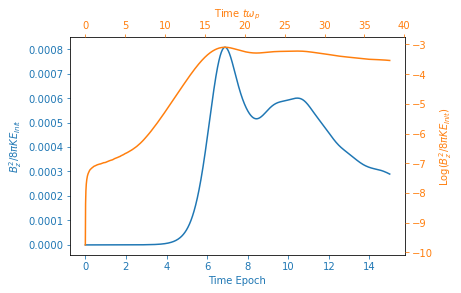}
    \end{subfigure}
    \vskip\baselineskip
    \begin{subfigure}
        \centering 
        \includegraphics[scale=0.4]{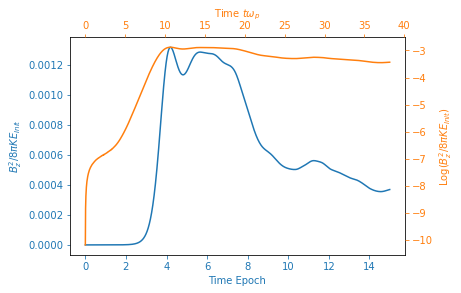}
    \end{subfigure}
    \begin{subfigure}
        \centering 
        \includegraphics[scale=0.4]{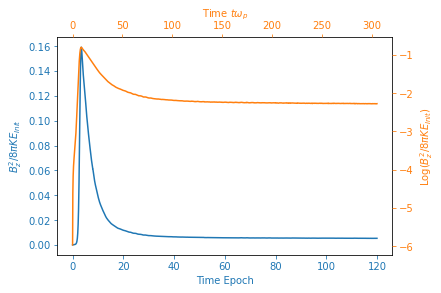}
    \end{subfigure}
    \caption{Panels showing the $|B_z|^2$ evolution of the four simulations run. With (a) S1, (b) S2, (c) S3, (d) S4. The units of $|B_z|$ are dimensionless - the code normalizes the numerical units by a term, $B_0 = (CC^2 \sqrt{\sigma_0}/COMP)\Delta x/\Delta t^2$. This is constructed of the Courant number, magnetization, skin depth in electrons per cell, grid length, and time interval, respectively. The magnetic field is then normalized by the total initial kinetic energy. Further explanation for these terms and their construction can be found in Appendix B. In the current numerical set-up, $|B_z|$ is the only non-zero magnetic field component. These plots show the sometimes drastic results the parameters can have on the WI evolution and if the secondary excited state becomes dominate at all. The curve in blue shows the magnetic energy evolution in linear linear scale while the orange curve shows the magnetic energy density in linear-log scale. The orange curve is used to calculate the slope between saturation (maximum) and beginning of filament construction (inflection before saturation). This slope represents the growth rate of the WI in simulation data.}
    \label{Fig:Bz_Evo_S1234}
\end{figure*}

\indent First, we present the evolution of the magnetic energy, $|B_z|^2 / 8\pi$. The current $2D$ numerical set-up recovers $|B_z|$ as the only non-zero component of the magnetic field. This field output is normalized within the simulation code in numerical units, creating a dimensionless output. The evolution of energy density is needed to deduce and characterize the evolution of the instabilities. This is also a method of studying how the different parameters affect the overall system. Each panel in (Fig. \ref{Fig:Bz_Evo_S1234}) shows the WI in a non-linear, relativistic regime as filaments grow due to electron bunching until saturation is reached (peak of curves). 

\begin{figure}
    \centering
        \includegraphics[scale=0.5]{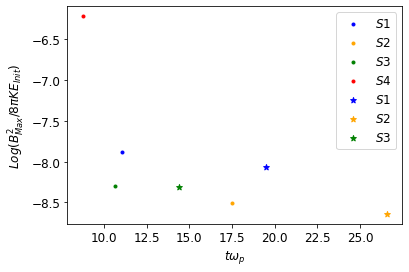}
    \caption{Comparison of $B_{z,Max}$ (circles) and $B_{z,Merge}$ (stars) as a result of the different simulations parameters is shown. The panel shows the relationship between the time of saturation ($t\omega_p$) and the magnetic field. Because there is no major merger event in $S4$, the data for that simulation is not present in the right panel.} 
    \label{Fig:Bmax_Comp_S1234}
\end{figure}

\indent Taking a higher particle density raises the energy contained in the filaments themselves. This results in a higher energy large filament merger event at late epochs and general lack of shallow tapering in the total magnetic energy. The half beam speed shows a longer time period until WI ignition, but a dissipation and filament merger on comparable time scales to the fiducial simulation. The $t\omega_p$ difference in events as a function of parameter can be seen in (Fig. \ref{Fig:Bmax_Comp_S1234}). The low skin depth run has a completely different timeline and energy signature. With the lower skin depth, the filaments were not allowed to be fully numerically resolved. Thus, the strong numerical dissipation prevents the system becoming chaotic enough to accelerate particles and produce radiation. Therefore, the WI dissipates in a toy system-like manner, without any further violent evolution. 

\begin{figure}[h]
    \centering
    \includegraphics[scale=0.5]{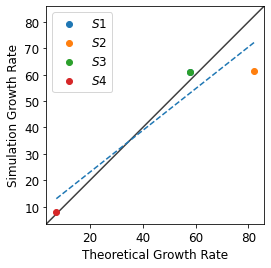}
    \caption{A comparison between the growth rate recovered by theory and the growth rate recovered by the simulation. The solid black line represents a $y=x$ trend, or a ratio of unity. The dashed line is a best fit line between the growth rate points. The closer the points are to a ratio of unity, the closer the simulation growth rate fits theory. The data point for $S1$ simulation is equal to the data point for $S3$ as their $\omega_p$ and $\gamma_{beam}$ terms are equal. See Table 1 in Appendix 2.}
    \label{Fig:Growth_Rate_Theory_Sim_Comp}
\end{figure}

\indent To ensure the simulations follows known theory, analysis was performed comparing the growth rate of the WI in theory with the growth rate seen in the simulations. The theoretical growth rate, $\Gamma_{WI}^{T}$, can be determined using $\Gamma_{WI}^{T} \equiv \omega_p/\sqrt{\gamma_{beam}}$. The $\omega_p$ term is the magnitude of both $\omega_{p,e^-}$ and $\omega_{p,e^+}$, due to the comparable nature of the pair plasma oscillations. The growth rate of the WI in the simulations, $\Gamma_{WI}^{S}$ is found by taking the point of inflection just before saturation  and the point of inflection as filament construction begins on the linear-log magnetic energy density evolution graphs and calculating the slope between these two points. This growth rate is then un-normalized to recover the raw values for comparison (Fig. \ref{Fig:Growth_Rate_Theory_Sim_Comp}). More information on the numerical values of this comparison can be seen in Appendix C.

\indent Examining these plots dictates the characteristic epochs that define the WI and subsequent merger activity. We define the following epochs for the WI with the numbers corresponding to the fiducial simulation: $1$ - thermal noise, $3$ - filament formation, $5$ - saturation, $8$ - filament merger. After the merger event, we no longer see signatures related to the WI. The following epochs are also defined as characteristic, but to Weibel dissipation: $11$ - dip in signature, $15$ - near thermal/relaxation. The characteristic epoch numbers and their associated evolutionary descriptor can be found in Table 2 in Appendix 3.

\begin{figure*}
    \centering
    \begin{subfigure}
        \centering
        \includegraphics[scale=0.5]{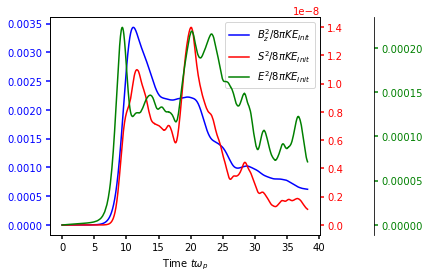}   
    \end{subfigure}
    \begin{subfigure} 
        \centering 
        \includegraphics[scale=0.5]{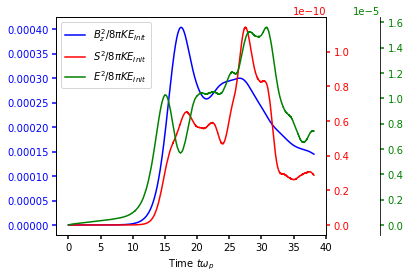}
    \end{subfigure}
    \vskip\baselineskip
    \begin{subfigure} 
        \centering 
        \includegraphics[scale=0.5]{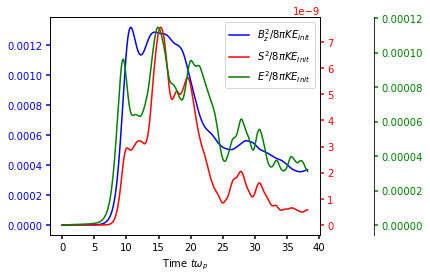}
    \end{subfigure}
    \begin{subfigure} 
        \centering 
        \includegraphics[scale=0.5]{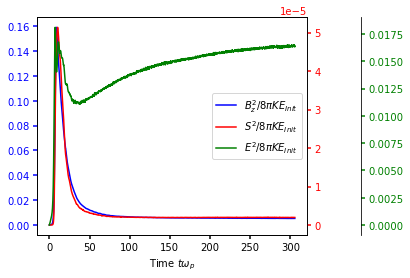}
    \end{subfigure}
    \caption{Simulations $S1$ (top left), $S2$ (top right), $S3$ (bottom left), and $S4$ (bottom right) evolutionary tracks with magnetic (blue), electric (green), and Poynting (red) energy density. The magnitudes are plotted relative to each other, so the different peaks are not numerically equal but are temporally equal. $S4$ shows heavy numerical noise and physical inaccuracies due to the numerical noise.} 
    \label{Fig:Triple_Field_Evo_S1234}
\end{figure*}

\indent Expanding from the magnetic field, we can study the relative magnitudes of the $|\vec{E}|^2$ and $|\vec{S}|^2$ evolutionary tracks (Fig. \ref{Fig:Triple_Field_Evo_S1234}). Here, $\vec{E} = \sqrt{E_x^2 + E_y^2}$ and $\vec{S} = \vec{E} \times \vec{B}$. The first feature to note is that the electric field is always less than the magnetic field with the exception of $S4$. In fact, $S4$ showcases very nonphysical attributes: greater electric field and a steadily growing electric field with no means of amplitude magnification. These phenomena are a product of inferior numerical resolution.

\indent For $S1$, $S2$, and $S3$, we see remarkable agreement that major filament merger coincides with the excitation of particle energies and a flux of energy, seen in the red and green lines peaking as the blue line reaches a local maxima representing the major filament merger ($\sim 20 t\omega_p$ for $S1$, $\sim 27 t\omega_p$ for $S2$, $\sim 17 t\omega_p$ for $S3$). This result demonstrates that the filament merger events generate strong electric fields capable of accelerating in situ particle populations. Between saturation and major merger, there is also a dip in Poynting flux, which represents the system attempting thermal relaxation as filaments drift without growing. These multi-track evolution plots also highlight other areas of energy excitation that could be regions of particle acceleration and radiation production.

\begin{figure}
    \centering
    \begin{subfigure}
        \centering
        \includegraphics[scale=0.3]{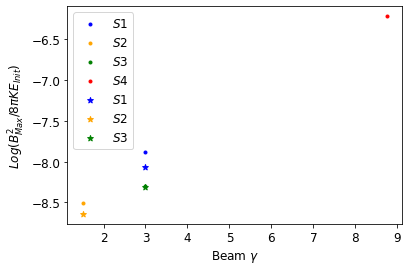}
    \end{subfigure}
    \begin{subfigure}
        \centering
        \includegraphics[scale=0.3]{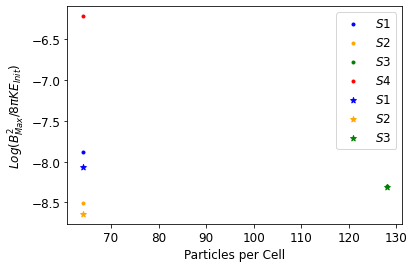}
    \end{subfigure}
    \begin{subfigure}
        \centering 
        \includegraphics[scale=0.3]{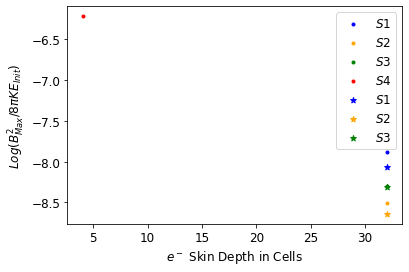}
    \end{subfigure}
    \caption{$B_{z,Max}$ (circles) and $B_{z,Merge}$ (magnetic field spike at filament merger) (stars) of all four simulations. $S4$ does not have a star point due to its lack of filament merger events. These plots show great agreement to the parameter relationships presented in theory.} 
    \label{Fig:Bz_Max_Parameters_S1234_Comp}
\end{figure}

\begin{figure}
    \centering
    \begin{subfigure}
        \centering
        \includegraphics[scale=0.25]{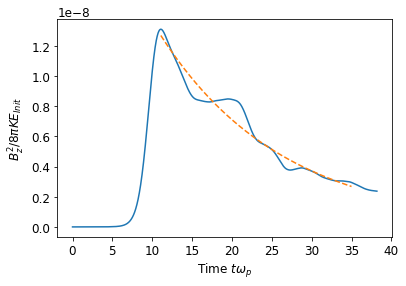}
    \end{subfigure}
    \hfill
    \begin{subfigure} 
        \centering 
        \includegraphics[scale=0.25]{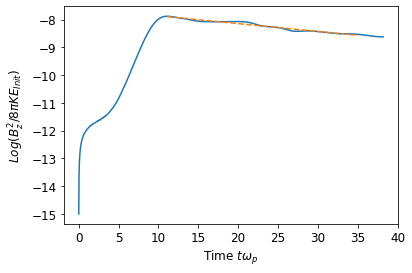}
    \end{subfigure}
    \vskip\baselineskip
    \begin{subfigure}
        \centering
        \includegraphics[scale=0.3]{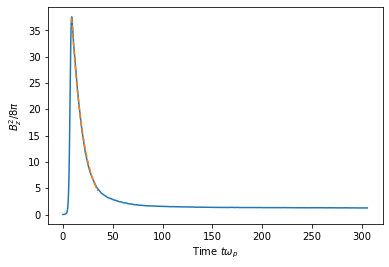}
    \end{subfigure}
    \hfill
    \begin{subfigure} 
        \centering 
        \includegraphics[scale=0.3]{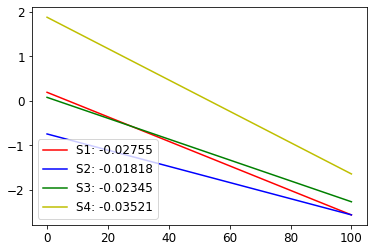}
    \end{subfigure}
    \caption{Linear-linear space magnetic field evolution and dissipation fitting of $S1$ (top right), the same plot but in linear log (top right), the dissipation fitting in linear-linear of $S4$ (bottom left), and the comparison of dissipation slopes in linear log of all four simulations (bottom right). The relationship between parameters, dissipation, and Weibel dissipation shown is complex but sensitive.}
    \label{Fig:S1234_Lin_Bz_Dissipation}
\end{figure}

\indent In an examination of some of the properties of the simulation parameters, we see these relationships reinforced. Through the evolutionary tracks, the parameter relationships shown in theory also show in the data (Fig.\ref{Fig:Bz_Max_Parameters_S1234_Comp}). The WI system evolution rate scales with the beam velocity, its overall power scales with the particles per cell, and its filament growth scales with skin depth. What is not shown by theory and we present here is the relationship the evolution has with the major filament mergers and how the dissipation of the WI is affected. (Fig. \ref{Fig:S1234_Lin_Bz_Dissipation}) shows how the dissipation rates were fitted in linear-log space and the respective slopes plotted in the same linear log space. We show here that the dissipation is greatest at lower skin depths consistent with strong numerical dissipation. Compared to $S1$ (red line), the lower beam velocity of $S2$ (blue line) not only shows a shallow gradient, but an overall lower dissipation, taken from the lower amount of energy from the beams. $S3$ (green line) shows a slightly greater amount of magnitude (more particles per cell), but shallow gradient compared to $S1$. These results show a complex relationship between parameters, dissipation, and filament interaction.

\indent We show the WI begins in a linear phase, whose time scale is determined by the beam velocity $\gamma_{beam}$. This is accompanied by an exponential growth of $B_z$ and the formation of current filaments seen in the magnetic field maps. The energy contained in these filaments is dictated by the particle number density present in the simulation. The growth rate of these filaments scales the skin depth of the system as presented in with theory. As shown, the skin depth of the system is the guiding parameter for latitudinal scales and both beam-parallel and -perpendicular filament numbers. This then heavily influences the remainder of the Weibel evolution. With filaments covering the simulation box, the instability system enters the saturation frame. We see the time of saturation scaling with $v_{beam}$, the power of the saturation scaling with $n_{PPC}$ and the dissipation of the instability from saturation scaling with electron skin depth $d_e = c/\omega_{p,e^-}$. 

\subsection{Physical Fields}
\begin{figure}[h]
    \centering
    \begin{subfigure} 
        \centering 
        \includegraphics[scale=0.3]{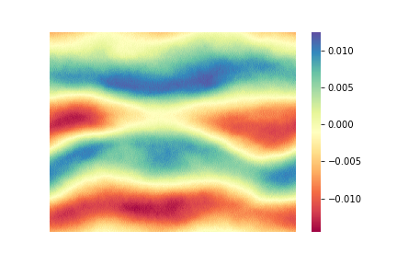}
    \end{subfigure}
    \begin{subfigure} 
        \centering 
        \includegraphics[scale=0.3]{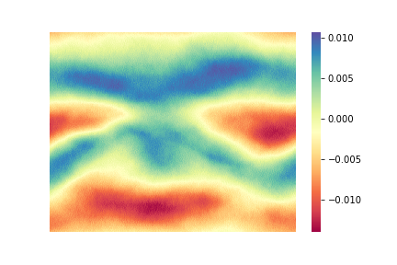}
    \end{subfigure}
    \caption{Snapshots $2150$ (left) and $2300$ (right) for $S1$ show the filament merger between the WI filaments (two blue filaments). This completes the WI evolution and the system then evolves into a disrupted dissipation due to the longitudinal electromagnetic and the latitudinal electrostatic wave modes. This event is shown in the secondary peak of the magnetic energy evolution when the skin depth is high enough for filament growth.}
    \label{Fig:S1_Filament_Merge_Snaps}
\end{figure}

\indent The focal point of the physical field analysis was studying the behavior of the $B_z$ field. The major observation seen in three of the four simulations was the presence of large merger events that coincided with secondary peaks in the magnetic energy evolution plots. This secondary peak is similar to the re-amplification of post-saturated WI shown described in Kumar et. al.\cite{KMK}.  These event occur between the two large filaments in the system (Fig. \ref{Fig:S1_Filament_Merge_Snaps}). 

\begin{figure}[h]
    \centering
    \includegraphics[scale=0.5]{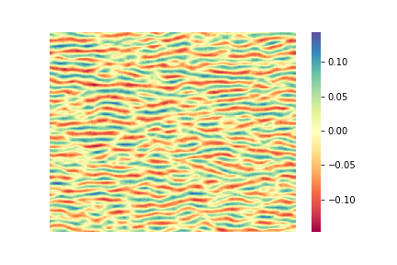}
    \caption{Snapshot $140$ of $S4$. With the low skin depth, the filaments are not allowed to grow to adequate sizes for violent filament mergers. This causes them to smoothly dissipate out and instability system to relax to thermal noise.}
    \label{Fig:S4_Skin_Depth_Snaps}
\end{figure}

\indent When the magnetic field for the lowered skin depth test ($S4$) was examined, the field showed an exponential growth in the number of filaments within the system. For the previous three tests, there was a ``countable" number of filaments (Figs. \ref{Fig:S1_Filament_Merge_Snaps}) present within the box. Now, filaments were fully populating the grid with little space in between (Fig. \ref{Fig:S4_Skin_Depth_Snaps}). Comparing the field plots with the field strength evolution shows that as the evolution hits the peak (saturation), the filaments begin to dissipate and smooth out without merging, again consistent with high numerical dissipation.

\begin{figure}
    \centering
    \begin{subfigure}
        \centering
        \includegraphics[scale=0.3]{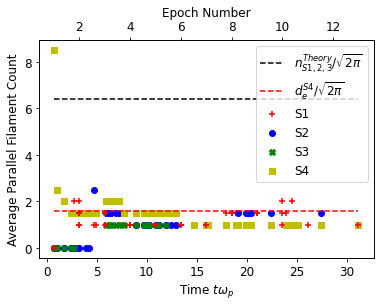}
    \end{subfigure}
    \hfill
    \begin{subfigure} 
        \centering 
        \includegraphics[scale=0.3]{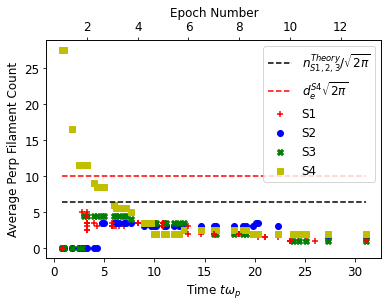}
    \end{subfigure}
    \vskip\baselineskip
    \begin{subfigure}
        \centering
        \includegraphics[scale=0.3]{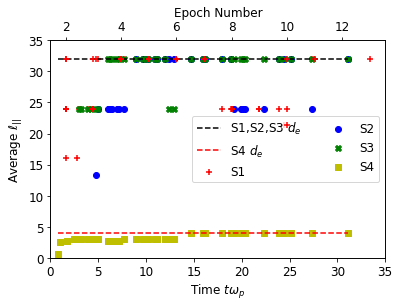}
    \end{subfigure}
    \hfill
    \begin{subfigure} 
        \centering 
        \includegraphics[scale=0.3]{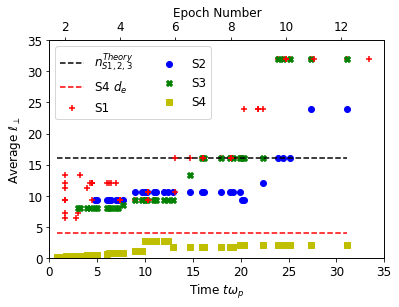}
    \end{subfigure}
    \caption{Examination of the average number of beam-parallel filaments (top left) and beam-perpendicular filaments (top right) and the average size or scale of beam-parallel (bottom left) and beam-perpendicular (bottom right) filaments. The black line shows the theoretical estimate for simulations $S1$, $S2$, and $S3$ who all share the same skin depth. The red line represents $S4$.}
    \label{Fig:S1234_Filament_Counting}
\end{figure}

\indent Examining the filament count and scales, we show an agreement with theory during the early stages of the WI. In the simulation run, the number of filaments in the box during the early epochs (pre-saturation) is on the order of the number of skin lengths that fit in the box side within a factor of $\sqrt{2\pi}$. We show in (Fig. \ref{Fig:S1234_Filament_Counting}) that the number of filaments and scale of filaments aligns with either the skin depth of the system of the theoretical count within a factor of $\sqrt{2\pi}$. A small uncertainty may be present on what was counted as a filament, as the interchange of filaments leaves magnetic bubble like structures across the box that are not quite filaments but not quite other structures \cite{21}. Looking at the top left panel of (Fig. \ref{Fig:S1234_Filament_Counting}), we see every simulation agreeing with the theoretical value for $S1$, $S2$, and $S3$. This is due to the unbounded beam parallel length of the filaments, which in turn stretch the entire length of the box, creating a single filament in the majority of cases. If the box stretched much further, the smaller filaments of $S4$ would eventually close and another would form, creating $n > 1$ filament in the beam parallel axis. We see great $S4$ agreement with the theory in the bottom left panel of (Fig. \ref{Fig:S1234_Filament_Counting}), as this system does not evolve into a post-saturation state due to filament mergers. This post-saturation state can be seen in the perpendicular filament lengths in the bottom right panel of (Fig. \ref{Fig:S1234_Filament_Counting}). With filaments merging on the perpendicular axis, the counts and lengths deviate from the theory (which does not account for merging), while the parallel values stay near the theoretical values.

\subsection{Dispersion Relations}
\indent Now we turn to the studies of the dispersion causes of plasma eigenmodes that are present in our simulation box at different epochs during our simulations. We perform the FFT analysis of the electric and magnetic field present in the box at each epoch: $B_z(x,y,t) -> \Bar{B}_z(k_x,k_y,t)$ and similarly for other fields. The FFT of the field is taken along with the corresponding $\omega$ and $k$ frequencies, the values are normalized, and then the data is plotted. What follows is a dispersion analysis of the WI in the fiducial simulation ($S1$).

\begin{figure} 
    \centering
    \begin{subfigure} 
        \centering 
        \includegraphics[scale=0.3]{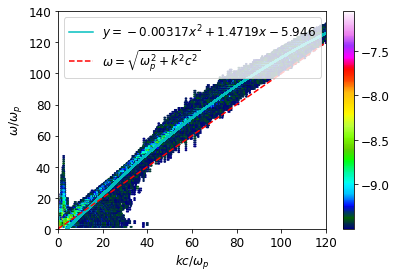}
    \end{subfigure}
    \hfill
    \begin{subfigure} 
        \centering 
        \includegraphics[scale=0.3]{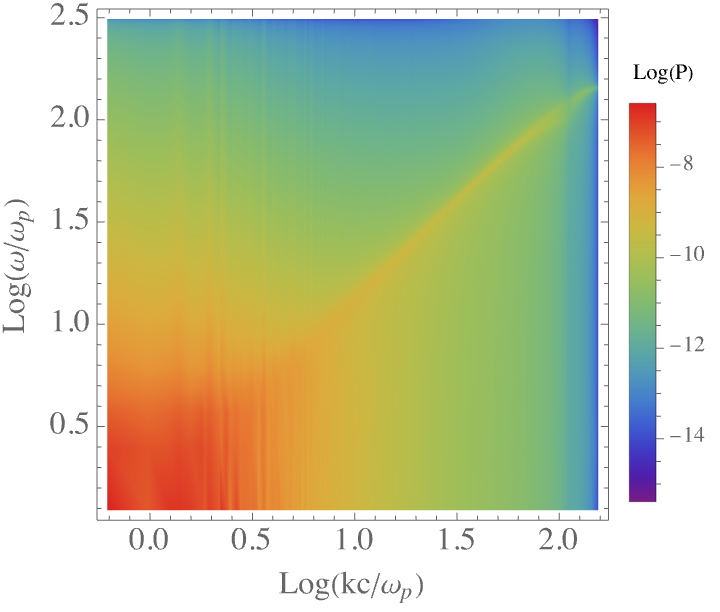}
    \end{subfigure}
    \caption{A depiction of the $B_z$ field dispersion relation at Epoch 3 with the dispersion function fit over the main dispersion mode, an isotropic electromagnetic wave. Epoch $5$ (saturation) is shown on the right in a traditional dispersion plot. These panel showcase the tapering exponential of the electromagnetic mode and the oscillation at the base of the curve. Not shown, epochs $1$, $8$ display very similar behavior.} 
    \label{Fig:Disp_Rel_S1_Bz_CEs}
\end{figure}

\begin{figure}
    \centering 
    \includegraphics[scale=0.5]{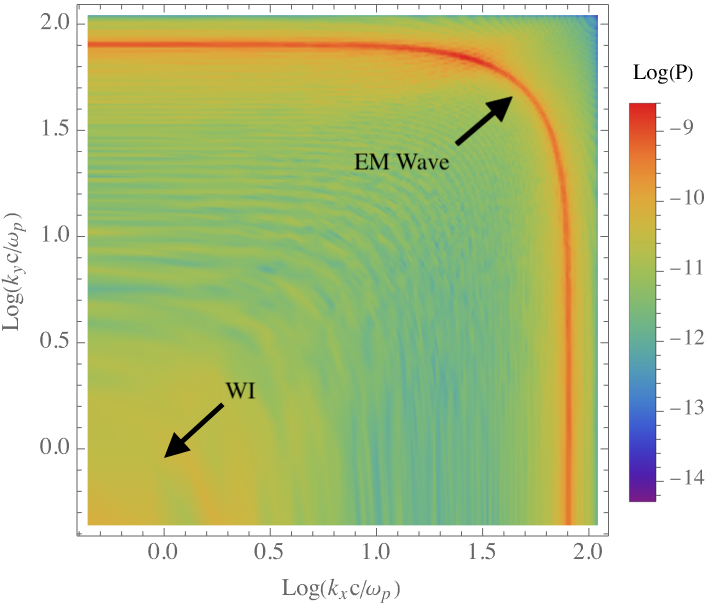}
    \caption{$B_z$ dispersion expansion for epoch $8$ for $\omega = 86.38$\footnote{This $\omega$ value corresponds to the frequency index in the epoch of $256$ snapshots.}. The high amplitude lateral bar is shown to dissipate as the WI evolves while the wave mode stays constant in amplitude in general. The labels on the bottom right panel show the different modes present in the system.} 
    \label{Fig:Dis_Rel_S1_Bz_Snaps}
\end{figure}

\begin{figure}
    \centering
    \includegraphics[scale=0.5]{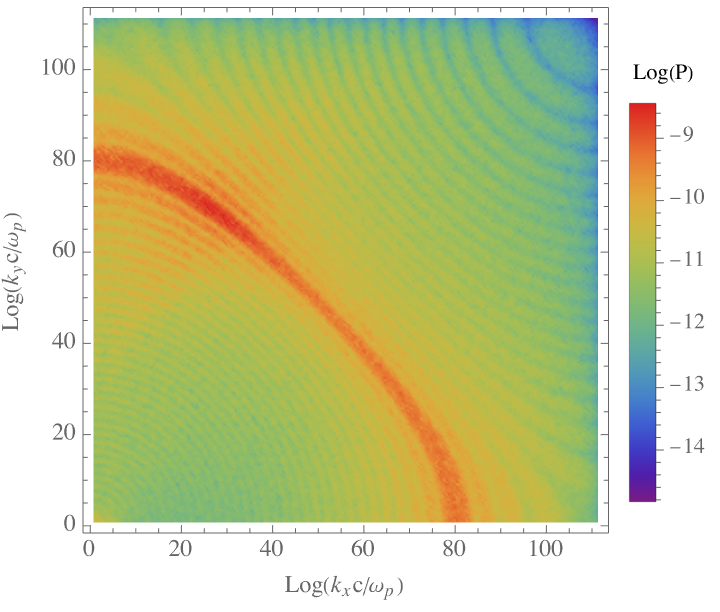}   
    \caption{A linear-linear dispersion of $B_z$ field in epoch $8$ ($\omega = 86.38$) depicting the isotopic electromagnetic wave propagating through k-space.}
    \label{Fig:Disp_Rel_Lin_Single}
\end{figure}

\indent First, consider magnetic field. We see a steadily tapering exponential in the total dispersion plot of the WI (Fig. \ref{Fig:Disp_Rel_S1_Bz_CEs}).  In the dispersion expansion, shown in (Fig. \ref{Fig:Dis_Rel_S1_Bz_Snaps}), an electromagnetic wave mode is shown propagating through the $k_x k_y$ box with the highest amplitudes on the high $k_y$ regime and tapering amplitude intensity as the wave propagates through the low $k_y$ values. The perfect circular arch represents isotopically propagating electromagnetic waves (Fig.  \ref{Fig:Disp_Rel_Lin_Single}). Another mode seen in the $k_x-k_y$ graphs is at the smallest $k$-values. This corresponds to the WI, which is present at large scales- at or above the plasma skin scales $c/\omega_p$ - and seen as the magnetic field filaments.  

\indent Fig. \ref{Fig:Disp_Rel_S1_Bz_CEs} shows the entire dispersion relation $\omega$ vs. $k$ for S1 for one the most important epochs ($5$ - saturation) where $k = \sqrt{k_x^2 + k_y^2}$. Here one can see the electromagnetic wave mode with the dispersion $\omega^2 = \omega_p^2 + k^2c^2$ and a low-$\omega$, low-$k$ mode corresponding to the slowly evolving large scale Weibel filaments.

\begin{figure}
    \centering 
    \includegraphics[scale=0.5]{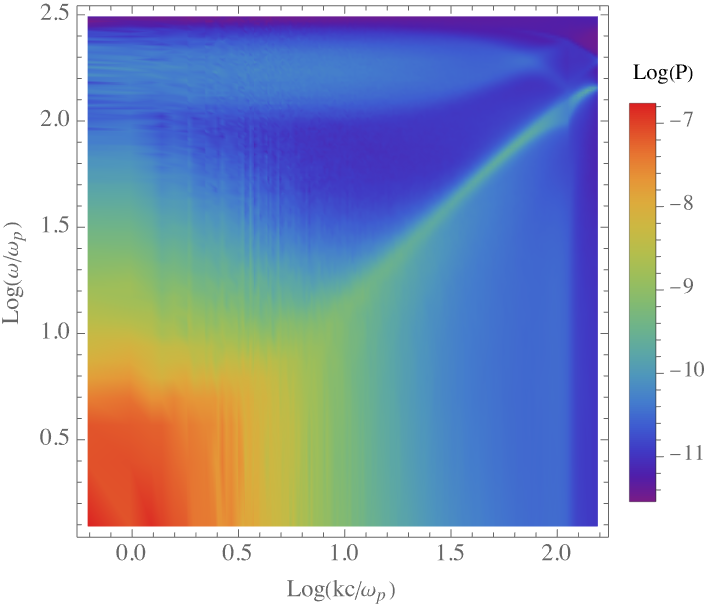}
    \caption{Depiction of the dual wave mode behavior of the total $|\Vec{E}|$ field. As the WI progresses through its evolution, the thermal noise epoch (not shown) displays a single, well defined wave. Filament ignition sees two well defined waves, the isotropic exponential curve and the faint linear mode at high $\omega$'s. Saturation and merger (not shown) display one well defined and a second faint wave mode.}
    \label{Fig:Disp_Rel_S1_E_Dual_Wave}
\end{figure}

\indent Second, consider the total $E$ field, $|\Vec{E}| = \sqrt{E_x^2 + E_y^2}$. We see in (Fig. \ref{Fig:Disp_Rel_S1_E_Dual_Wave}) a more complex behavior present. Throughout each epoch, an isotropically propagating electromagnetic wave (similar to $B_z$ maintains a maximum $k_y$ value stretched over all available $k_x$ values. At larger omega values, unlike $B_z$, a secondary wave is seen propagating through the box anchored at high $k_x$ values (spanning a up to the highest $k_y$ values). This is a purely numerical mode. It is a numerical artifact of the mirroring technique utilized in the FFT procedure. Because the FFT over $t \ \in \ [0, 256]$ yields $\omega \ \in \ [-128, 128]$, everything after $\omega = 128$ is not only the reverse of $\omega \ \in \ [0, 128]$, but is subject to numerical artifacts. During the last two characteristic epochs, the electromagnetic wave propagates again but it does not give way to a numerical mirror mode. Instead, a second isotropic EM wave propagates from the high-$k_x$ high-$k_y$ area of the box. With both physical and numerical behaviors present, it is necessary to analyze the dispersion expansions of the component ($E_x$ and $E_y$) fields as well.

\begin{figure}
    \centering
    \begin{subfigure} 
        \centering 
        \includegraphics[scale=0.3]{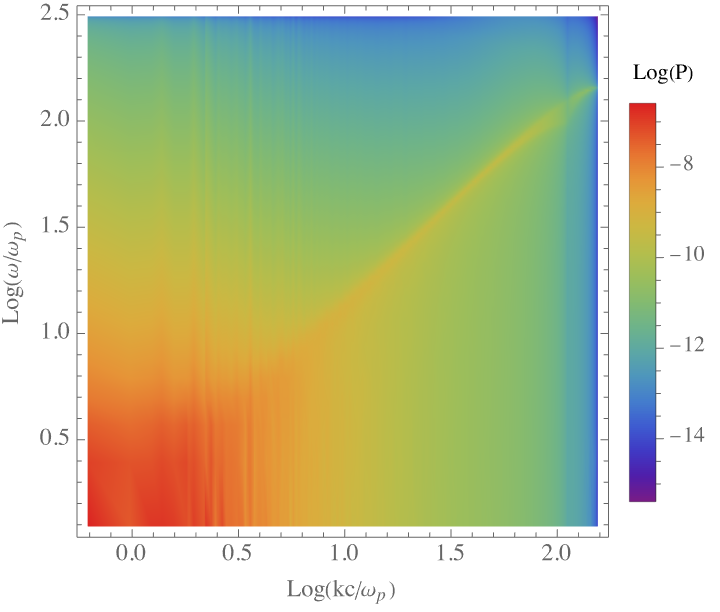}
    \end{subfigure}
    \hfill
    \begin{subfigure} 
        \centering 
        \includegraphics[scale=0.3]{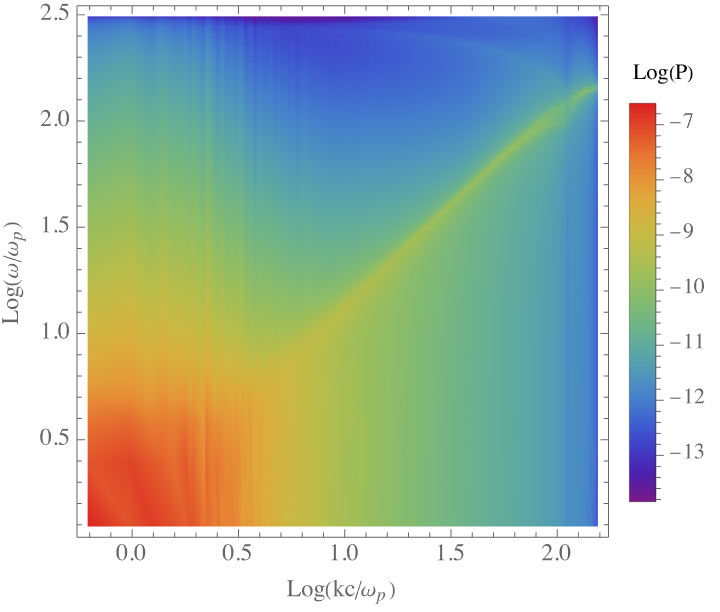}
    \end{subfigure}
    \vskip\baselineskip
    \begin{subfigure} 
        \centering 
    \includegraphics[scale=0.3]{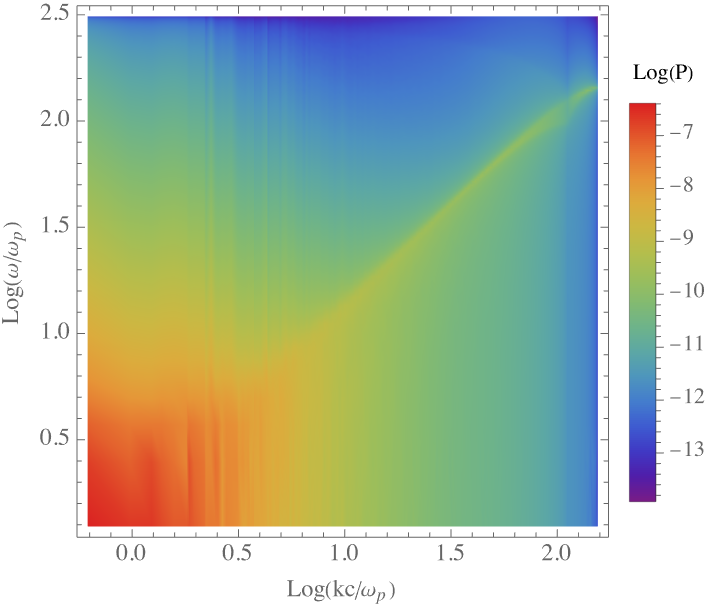}
    \end{subfigure}
    \hfill
    \begin{subfigure} 
        \centering
        \includegraphics[scale=0.3]{Disp_E5E.png}
    \end{subfigure}
    \caption{The total dispersion relation for epoch $5$ for the $B_z$ (top left), $E_x$ (top right), $E_y$ (bottom left), an $E$ (bottom right) fields. The fields perpendicular to the beam motion show great similarities during the most meaningful epochs of the WI evolution. The $E_x$ field shows the beam instability in the form of extra structures. Hook like structures can be seen at the tip of the dispersion curve which can be related to beam motion present in the system. $|\vec{E}|$ depicts the sum of the component fields, with the small oscillation at the base of the curve seen in the $E_y$ field. }
    \label{Fig:Disp_Rela_S1_E_CEs}
\end{figure}

\begin{figure}
\centering
    \begin{subfigure} 
        \centering 
        \includegraphics[scale=0.3]{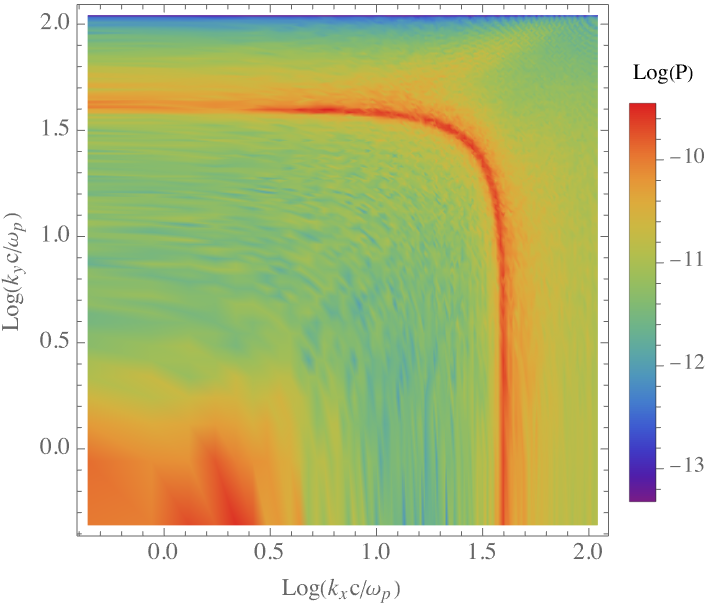}
    \end{subfigure}
    \hfill
    \begin{subfigure} 
        \centering 
        \includegraphics[scale=0.3]{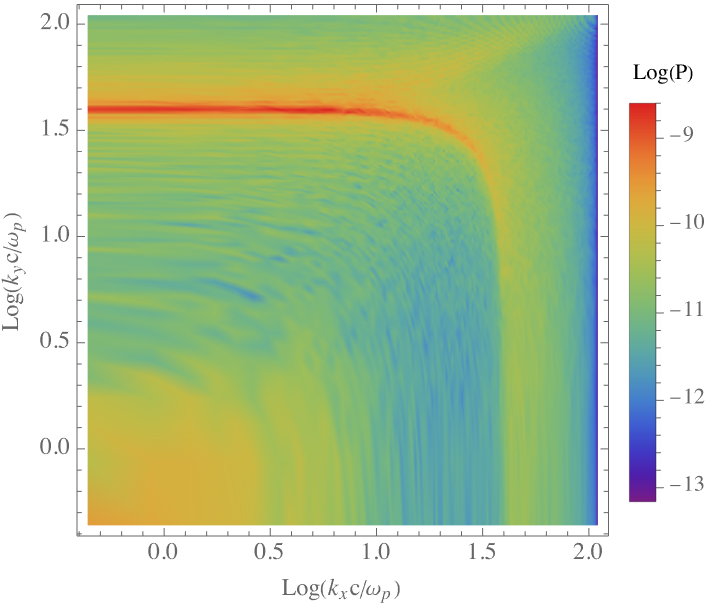}
    \end{subfigure}
    \caption{The $k_x-k_y$ plots for $\omega = 49.36$ for the same epoch but for the $E_x$ fields (left). As these fields are parallel to the beam motion, we see a beam instability present in the dispersion expansion in the form of reverberations behind the wave modes. For $E_y$ (right), we again see the dissipating high amplitude bar and a well defined wave mode.}
    \label{Fig:Disp_Rela_S1_Ey_Snaps}
\end{figure}

\indent In the $E_y$ field (Fig. \ref{Fig:Disp_Rela_S1_E_CEs} bottom left) we see again a single, slightly anisotropic - in amplitude - propagating wave that depicts a  more defined structure in the high-$k_y$ values (Fig. \ref{Fig:Disp_Rela_S1_Ey_Snaps}). This wave does not lead to a scanning mode. The $E_x$ field, on the other hand, showcases more complex behavior over the WI lifetime. These do not evolve in to any scanning modes in later high-$\omega$ values. Instead we see a "reverberation'' pattern trailing the propagating wave in low-$\omega$ (Fig. \ref{Fig:Disp_Rela_S1_E_CEs}). The reverberation pattern shows a wake of isotropic waves trailing the lead propagating wave. This reverberation drastically decreases in amplitude over the different epochs, but is present in each epoch. The lack of extra modes present in the $E_y$ field when compared to the $E_x$ field is expected when noting the beams propagate in the x-direction and that the Langmuir waves generated in the system are longitudinal.

\indent Looking at the total dispersion relations, we see similar behavior in the $B_z$ and $E_y$ (Fig. \ref{Fig:Disp_Rela_S1_E_CEs}) fields. They both show an isotropic dispersion relation with tapering at high-$k$ and high-$\omega$ values and oscillation of high amplitude at low-$k$ low-$\omega$. The $E_x$ field shows less oscillation at low-$k$ low-$\omega$, but does showcase a small structure at high-$k$ high-$\omega$ values (Fig. \ref{Fig:Disp_Rela_S1_Ey_Snaps}). This structure, which could be the reverberations seen in the expansions, is the the result of the field being parallel to the beaming direction. The $E$ field shows little to no oscillations at low-$k$ low-$\omega$. This manifests the absence of purely low-$k$ modes and quasi-neutrality of filaments.

\indent We see through the dispersion analysis that the dominate mode is an electromagnetic wave excited by the WI in the low $k-\omega$ regime in all fields. The $E_x$ field dispersion presents many different and complex structures to discuss. The hooks at the top of the dispersion curve are thought to be a electrostatic instability excited by the beams itself. This electrostatic beam instability is also the cause of the reverberation pattern seen in (Fig. \ref{Fig:Dis_Rel_S1_Bz_Snaps}). The rapid dissipation over the epochs is a result of phase mixing or Landau damping. 

\indent The two instabilities available to generate an electrostatic seen in the $E_x$ field are the Buneman Instability (BI) and the Two-Stream Instability (TSI). The BI is caused by particle drift velocity exceeding the sound speed in the plasma and exciting acoustic waves. These acoustic waves are heavily suppressed by Landau damping. This rules out the BI.

\indent The TSI is a more generic form of electrostatic instability. This will be nearly equivalent to the inverse of Landau damping in a hot plasma, like the ones present in the simulations. If the simulations possessed a bulk plasma and a weak beam, the instability would follow the signatures of the beam-plasma instability. The simulations run in this work, instead, have two strong beams (one in place of a bulk plasma). This leads to the conclusion that the reverberations are caused by a longitudinally excited wave from the electrostatic two-stream instability.

\indent This longitudinal wave could be the Oblique instability (OI), whose signature is also a checkerboard mode from two perpendicular wave modes. The latitudinal electromagnetic perturbation is excited by the WI and the longitudinal perturbation is excited by the electrostatic two-stream instability from the strong beams. The coexistence of these two modes in the WI lifetime, and the large filament merger event present in a system of countable filaments may create an environment suitable for the OI), but more analysis must be done to confirm the origins and presence of the OI within the system. Our findings will be presented in subsequent publications. 

\section{Conclusions}
\indent In this paper we have shown that the Weibel instability after saturation of filamentation relaxes further via filament mergers. The analysis of the fields shows that the number of filaments (major and minor) in the system is dictated by the skin depth of the simulation. In addition, the number density of particles dictate how strong the filaments are within the system, though the results argue it is only in relation to the post-saturation state, as the total system energies is comparable to that of the control. The velocity of which the beam is being propagated does have an effect on the system's total magnetic field energy, as one would expect. It also shifts the saturation point and therefore the Weibel timeline. The dispersion analysis showed a coupled electromagnetic wave propagating through the system at each epoch. The magnetic field density evolution of each simulation shows a similar evolutionary track with differences based on the parameters tested (beam propagation velocity, particle density, and skin depth).  When the filaments are sufficiently large, the time scale of the system is large enough for these filaments to merge under violent merger events. This rapidly accelerates the Weibel turbulence relaxation while the plasma is not yet relaxed, creating a highly magnetized, locally-unstable environment. This new environment allows the secondary excited state to become important in the system.  

\indent The importance of these findings is solid. The violent relaxation of ``post-saturated'' Weibel turbulence via filament mergers provides a vigorous mechanism for particle acceleration and collisionless plasma heating. Although the Weibel instability garners only ten percent of the energy budget of the system, there is still a question of where the rest of the energy is allocated. There is also the question of radiation production. The Weibel instability theoretically hosts and produces Jitter radiation. The Weibel turbulence and mergers create a new mechanism for this radiation production to be modeled and examined. There is also the question of the violent merger event producing this Jitter radiation. Finally, violent filament merger events happening within the plasma bring forth the means to propose a solution to the injection problem in collisionless Weibel-mediated shocks. These events could serve as the accelerators for particles to breach thermal velocities and be active participants in diffusive shock acceleration. The aforementioned questions are currently being investigated and the results will be presented in subsequent publications. 



%
%

%

\begin{acknowledgments}
This study is partly supported by PHY-2010109. This research made use of "TRISTAN-MP v2" particle-in-cell code. The author would like to thank Dr. Philippov for his discussion, comments, and suggestions and Dr. Medvedev for his guidance and role of graduate advisor for this work and subsequent works.
\end{acknowledgments}

\appendix

\section{\label{sec:level1}Simulation Parameters}
\indent Below is a table of parameters for the TRISTAN-MP simulation software corresponding to the four simulations ran in this and subsequent studies. The following section details how various parameters and characteristics of the system are derived and calculated\cite{26}. Parameters can be calculated using the following expressions found in the TRISTAN-MP wiki\cite{27}. For more detailed information please see the links to the wiki found in the references.

\indent A brief note about normalization. From the code developers: ``remember the parameters $CC$, $COMP$, and $PPC$ are purely numerical and thus arbitrary from a physical standpoint.'' This entails that simulation data was normalized within the arbitrary system of computational units defined by the parameters within the input system. This makes any inconsistencies, that from a physical standpoint would be incorrect ($\omega_{p,e^-}$ for different $PPC$ values being equal) are therefore legal within the computational system. Because all of the simulations were normalized within the same arbitrary unit system (as defined above) and compared to each other (not to any physical reference frame), the normalization performed and the results derived from the data remain valid.

\subsection{\label{sec:level2}Normalization in Code Units}
\indent Within the input file for the simulation runs, the parameter $c$ is defined as ``velocity of light in comp. units (this defines timestep).'' The is also an input parameter $Corr.$ which is the ``correction for the speed of light.'' In a script for field calculation in the code, under a ``full timestep'' do-loop, a constant is defined as 
\begin{equation}
    Const = Corr \times c,
\end{equation}
with a ``half-timestep'' including a factor of $0.5$. With the input files all defining $c = 0.45$ and $Corr. = 1.025$, this constant term for a full timestep is given as $0.46125$. \\
\indent The ``measure of all spatial and temporal units in cell size and timesteps'' is denoted as the variables $\Delta x$ and $\Delta t$, respectively. The code is also normalized in a way that $\Delta x = \Delta t$. This, combined with the expression $c = CC \frac{\Delta x}{\Delta t}$ gives way to the assumption that $\Delta x = \Delta t \equiv 1$ and $CC = 0.45$. This value of $\Delta x$ defines the simulation particles per grid space as $PPC$, following the expression $n_{ppc} = PPC\Delta x^{-3}$. Nowhere in the input files or user files was the length, sampling, or space altered, allowing for the assumption that a grid space and a timestep are both a single unit interval. 

\indent From the definitions of a single timestep and single grid space, we may move onto to the plasma parameters themselves. In the parameter output file, a value defined as $Time$ is given. This can seen in, taken from the wiki's ``Code Unit Calculator''
\begin{equation}
    \omega_p^{-1} = Time \ \Delta t.
\end{equation}
Note: while the wiki says $\omega_p$, in our system this is strictly $\omega_{e^-}$, as we can not ignore the positrons with their similar frequencies.
\begin{equation}
    \omega_p = \sqrt{\omega_{e^-}^2 + \omega_{e^+}^2} = \sqrt{2}\omega_{e^-}.
\end{equation}
This defines, in the simulation's units, the length of a single timestep in units of $\omega_p$ ($\Delta t = \frac{1}{\sqrt{2}Time}$). This also defines the $\omega$ frequency data values' normalization constant (excluding the FFT normalization) as $\omega_p$.

\indent From $\omega_p$ and the temporal frequency normalization we may calculate the spatial frequency normalization and the skin depth in simulation units. The input files defines $COMP$ as ``electron skin depth in cells.'' With the documentation noting the fiducial skin depth value (to be discussed later) as $d_{e^-}^0 = COMP\Delta x$. This expression also serves as the definition of cell size in terms of $c/\omega_p$ as the value $1/COMP$. But to normalize the data properly, the skin depth must be calculated in the same simulation units as $\omega_p$. This can be found by the customary definition using our previous normalization values
\begin{equation}
    d_p = \frac{c}{\omega_p} = \frac{c}{\sqrt{\omega_{e^-}^2 + \omega_{e^+}^2}} = \frac{0.45}{\sqrt{2}\omega_{e^-}} = \frac{(0.45)Time}{\sqrt{2}}.
\end{equation}

\begin{figure}
    \centering
    \begin{subfigure}
        \centering
        \includegraphics[scale=0.3]{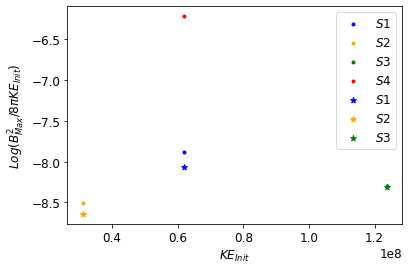}
        \label{fig:mean and std of net14}
    \end{subfigure}
    \begin{subfigure}   
        \centering 
        \includegraphics[scale=0.3]{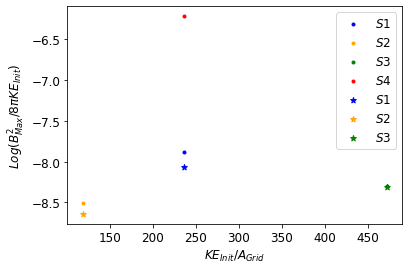}
        \label{fig:mean and std of net34}
    \end{subfigure}
    \caption{$B_{z,Max}$ (circles) and $B_{z,merge}$ (magnetic field spike at filament merger) (stars) of all four simulations. $S4$ does not have a star point due to its lack of filament merger events. The relationships between the initial kinetic energy (or energy density) and the magnetic field peaks are shown.}
    \label{Fig:Appendix_KE_Init_Den}
\end{figure}

\begin{figure}
    \centering
    \begin{subfigure}
        \centering
        \includegraphics[scale=0.28]{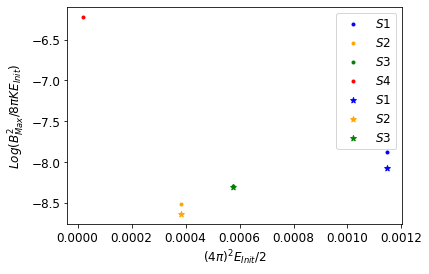}
    \end{subfigure}
    \hfill
    \begin{subfigure}
        \centering 
        \includegraphics[scale=0.28]{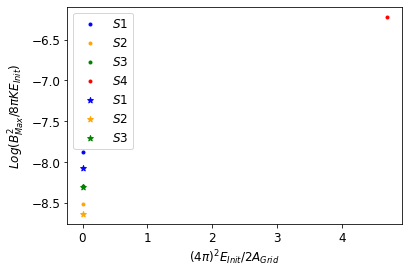}
    \end{subfigure}
    \caption{$B_{z,Max}$ (circles) and $B_{z,merge}$ (magnetic field spike at filament merger) (stars) of all four simulations. $S4$ does not have a star point due to its lack of filament merger events. The relationships between the initial potential energy (or energy density) and the magnetic field peaks are shown.}
    \label{Fig:Appendix_EPE_Init_Den}
\end{figure}

\indent The final normalization term to be computed is the energy normalization for frequency amplitudes.. This is derived from the first snap shot output itself and isn't directly associated with input parameters like spacial and temporal frequencies. Each of these normalization terms used is the energy density ($\frac{Energy}{A_{Grid}}$) and not the total energy itself. The first value to be calculated is the potential energy contained in the electric field. While this isn't directly used in the data, as the value is much smaller than that of the kinetic energy and therefore has no effect, it is important still to note. The total energy potential in the electric fields (Fig. \ref{Fig:Appendix_EPE_Init_Den}) can be found by calculating the total electric field thus
\begin{multline}
    U_{Electric, Init} = \frac{(\sqrt{(4\pi E_x)^2 + (4\pi E_y)^2})^2}{2} = \frac{(4\pi)^2E}{2}.
\end{multline}
Each value above is summed over each grid point to recover a total potential energy and then divided by the area for energy density (Fig. \ref{Fig:Appendix_EPE_Init_Den})
\begin{equation}
    U_{Electric, Init}^{Density} = \frac{16\pi^2E}{2A_{Grid}}.
\end{equation}
\indent Next, and more importantly, is the kinetic energy density (Fig. \ref{Fig:Appendix_KE_Init_Den}). For this calculation, the initial output of $i$ amount of particles energy ($\gamma$ in output) is summed over.
\begin{equation}
    K_{Particle, Init} = \sum_{n = 1}^i \gamma_i^{e^-} + \sum_{n = 1}^i \gamma_i^{e^+},
\end{equation}
then the density (Fig. \ref{Fig:Appendix_KE_Init_Den})
\begin{equation}
    K_{Particle, Init}^{Density} = \frac{K_{Particle, Init}}{A_{Grid}}.
\end{equation}

\subsection{\label{sec:level2}CGS Unit Conversion}
\begin{figure}
        \centering
        \begin{subfigure}
            \centering
            \includegraphics[scale=0.3]{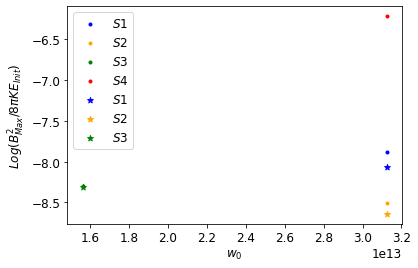}
        \end{subfigure}
        \caption{$B_{z,Max}$ (circles) and $B_{z,merge}$ (magnetic field spike at filament merger) (stars) of all four simulations. $S4$ does not have a star point due to its lack of filament merger events. The relationships between the number of real particles per macroparticles and the magnetic field peaks are shown.}
        \label{Fig:Appendix_Macroparticles}
\end{figure}

\indent As stated prior, the computational units hold no relation to real world units or physical systems. This, combined with the exclusion of some parameters in the input file, make the comparison of our systems to CGS unit based systems tricky. From the documentation, only one parameter must be defined in CGS units for the rest to then be converted and defined. We choose our first CGS definition as $\Delta x = 1 cm$. Recall the relation
\begin{equation}
    c = CC \frac{\Delta x}{\Delta t},
\end{equation}
with $CC = 0.45$ (the first fundamental dimensionless quantity) from the code. This recovers $\Delta t = 1.501 \times 10^{-11} s$. \\
\indent The second fundamental dimensionless quantity is the $PPC$ input parameter. The relation
\begin{equation}
    n_0 = w_0PPC\Delta x^{-3},
\end{equation}
allows the fiducial number density ($n_0$), fiducial macro-to-real particle ratio ($w_0$),
and particles per cell to all be related. Taking a number density of electrons as $10^{15} \frac{e^-}{cm^3}$, with positrons possessing the same value. This gives our ``real life'' system a total number density of $n_0 = 2 \times 10^{15}$. From the total particle density, the value of $w_0$ real particles represented by each simulation macroparticle can be found (Fig. \ref{Fig:Appendix_Macroparticles}).

\begin{figure}
        \centering
        \begin{subfigure}
            \centering
            \includegraphics[scale=0.3]{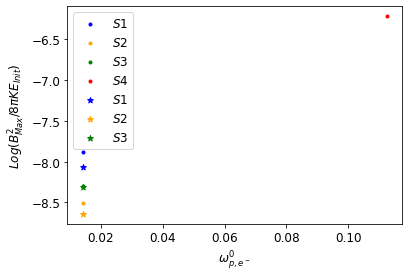}
        \end{subfigure}
        \hfill
        \begin{subfigure}
            \centering 
            \includegraphics[scale=0.3]{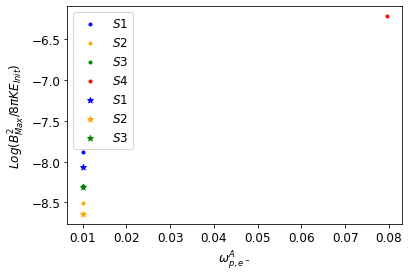}
        \end{subfigure}
        \caption{$B_{z,Max}$ (circles) and $B_{z,merge}$ (magnetic field spike at filament merger) (stars) of all four simulations. $S4$ does not have a star point due to its lack of filament merger events. The relationships between the fiducial plasma frequency and the magnetic field peaks are shown.}
         \label{Fig:BMax_omega_es_Appendix}
\end{figure}

\begin{figure}
        \centering
        \begin{subfigure}
            \centering
            \includegraphics[scale=0.3]{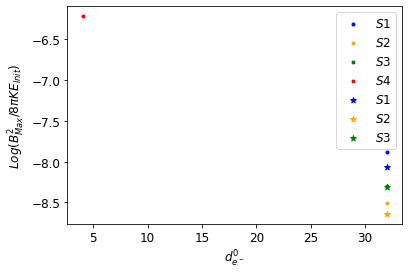}
        \end{subfigure}
        \hfill
        \begin{subfigure}
            \centering 
            \includegraphics[scale=0.3]{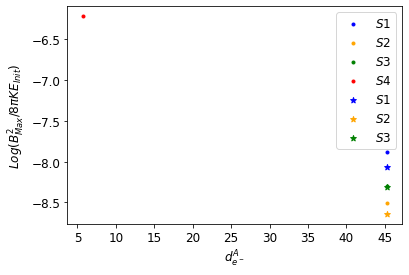}
        \end{subfigure}
        \caption{$B_{z,Max}$ (circles) and $B_{z,merge}$ (magnetic field spike at filament merger) (stars) of all four simulations. $S4$ does not have a star point due to its lack of filament merger events. The relationships between the fiducial skin depths and the magnetic field peaks are shown.}
        \label{Fig:BMax_d_es_Appendix}
\end{figure}

\indent The corresponding skin depth can now be expressed through though the last fundamental dimensionless parameter, $COMP$. From the input parameters, the fiducial skin depth value for electrons can be found via
\begin{equation}
    d_{e^-}^0 = COMP \Delta x.
\end{equation}
This expression introduces the ``fiducial term'' to the evaluation of unit based values. The documentation gives any value with a subscript of zero the fiducial label. These values are measured in the code units and can be used within the computational units system. This nomenclature is used to distinguish between the mentioned fiducial values in comp. units (change based on inputs), the formula of the value in CGS physics (unchanging), and the actual value in a physical framework with the input values used to reflect the simulations. For $\omega_{s}$ for particle species $s$ (which is the same in this analysis framework) we find the formula being the traditional
\begin{equation}
    \omega_{p,s} = \sqrt{\frac{4\pi n_s q_s^2}{m_s}},
\end{equation}
the fiducial value can be found through
\begin{equation}
    \omega_{p,e^-}^0 = \frac{CC}{COMP\Delta t},
\end{equation}
and finally the actual (comparative) value
\begin{equation}
    \omega_{p,s} =\frac{\omega_{p,e^-}^0 \sqrt{\hat{n}_s}|\hat{q}_s|}{\sqrt{\hat{m}_s}},
\end{equation}
shown in (Fig. \ref{Fig:BMax_omega_es_Appendix})
In these and the following equations, variables with a hat are defined by $\hat{x}_s = \frac{x_s}{x_0}$. For our purposes, the ratio is the same for each species (electron or positron), so these terms take the values: $\hat{n}_s = n_s/n_0 = 0.5$, $\hat{m}_s = m_s/m_e = 1$, and $\hat{q}_s = q_s/q_e = 1$.
For skin depth we have the following final three equations (Fig. \ref{Fig:BMax_d_es_Appendix}):
\begin{equation}
    d_S = \frac{c}{\omega_{p,s}},
\end{equation}
\begin{equation}
    d_{e^-}^0 = COMP \Delta x,
\end{equation}
and
\begin{equation}
    d_s = \frac{d_{e^-}^0\sqrt{\hat{m}_s}}{\sqrt{\hat{n}_s}|\hat{q}_s|}.
\end{equation}

\clearpage 
\onecolumngrid

\begin{table}[h]
\begin{center}
\caption{Parameters for the simulation code TRISTAN-MP.}
\begin{tabular}{ |p{5.0cm}||p{2.5cm}|p{3cm}|p{3cm}|p{3cm}|  }
 \hline
 \multicolumn{5}{|c|}{TRISTAN-MP Parameters} \\
 \hline
Simulation Parameter & S1 - Fiducial & S2 - Low $\gamma$ & S3 - High $PPC$ & S4 - Low $d_{e^-}$ \\
 \hline
 \multicolumn{5}{|c|}{Universal Parameters}\\
  \hline
 X Direction CPUs & \multicolumn{4}{|c|}{16}\\
 Y Direction CPUs & \multicolumn{4}{|c|}{4}\\
 X Direction Grid & \multicolumn{4}{|c|}{512}\\
 Y Direction Grid & \multicolumn{4}{|c|}{512}\\
 c (Comp. Units) & \multicolumn{4}{|c|}{.45}\\
 Correction to c & \multicolumn{4}{|c|}{1.025}\\
 Grid interval & \multicolumn{4}{|c|}{1}\\
 Timestep Interval & \multicolumn{4}{|c|}{1}\\
 Smoothing Filter Passes & \multicolumn{4}{|c|}{0}\\
 Magnetization Number & \multicolumn{4}{|c|}{0.00}\\
 Max Number of Particles & \multicolumn{4}{|c|}{$ 1 \times 10^9$}\\
 Delta $\gamma$: ($\frac{kT_i}{m_ic^2}$) & \multicolumn{4}{|c|}{0.2}\\
 Magnitization Number ($\sigma_0$) & \multicolumn{4}{|c|}{0}\\
 \hline
 \multicolumn{5}{|c|}{Simulation Input Parameters}\\
 \hline
 $\gamma_{Beam}$ & 3 & 1.5 & 3 & 3\\
 Particles per Cell & 64 & 64 & 128 & 64\\
 $e^-$ Skin Depth in Cell & 32 & 32 & 32 & 4\\
 \hline
 \multicolumn{5}{|c|}{Growth Rate}\\
 \hline
 Theoretical Rate ($\Gamma_{WI}^T$) & 58.07 & 82.12 & 58.07 & 7.25\\
 Simulation Rate ($\Gamma_{WI}^S$) & 61.06 & 61.43 & 61.06 & 7.94\\
 Growth Rate Ratios & 0.95 & 1.33 & 0.95 & 0.91\\
 \hline
 \multicolumn{5}{|c|}{Energy in Comp. Units}\\
 \hline
 Initial KE & 61884625.61 & 30941686.08 & 123767220.92 & 61884624.50 \\
 Initial KE Density & 236.07 & 118.03 & 472.13 & 236.07 \\
 Initial Electric PE & 0.0011 & 0.0003 & 0.0005 & $1.79 x 10^{-5}$ \\
 Initial Electric PE Density & $4.37 \times 10^{-9}$ & $1.45 \times 10^{-9}$ & $2.18 \times 10^{-9}$ & 4.69\\
 \hline
 \multicolumn{5}{|c|}{Magnetic Energy and Time in $\omega_p$}\\
 \hline
 $B_{Max}$ at Saturation & 0.81 & 0.09 & 0.62 & 37.55 \\
 $B_{Max}$ at Merger & 0.52 & 0.07 & 0.60 & N/A \\
 $\omega_p$ at $B_{Max}$ at Saturation & 11.07 & 17.52 & 10.67 & 8.75 \\
 $\omega_p$ at $B_{Max}$ at Merger & 19.49 & 26.56 & 14.40 & N/A \\
 \hline
 \multicolumn{5}{|c|}{Simulation Derived Parameters}\\
 \hline
 $\omega_{p}^{e^\pm}$ (Comp. Units) & 71.12 & 71.12 & 71.12 & 8.88\\
 $\omega_{p}$ (Comp. Units) & 100.58 & 100.58 & 100.58 & 12.57\\
 Skin Depth $d_{e^\pm} = \frac{c}{\omega_{p}^{e^\pm}}$ & 0.006 & 0.006  & 0.006 & 0.05 \\
 Skin Depth $d_{p} = \frac{c}{\omega_{p}^{p}}$ & 0.004 & 0.004 & 0.004 & 0.035 \\
 Step Size ($\omega_p^{-1})$ & 100.58 & 100.58 & 100.58 & 12.57 \\
 Cell Size ($c/\omega_p$) & 0.03125 & 0.03125 & 0.03125 & 0.25 \\
 Plasma React. Time $\frac{c}{d_{e^-} in \ Cells}$ & & & &\\
 \hline
 \multicolumn{5}{|c|}{TRISTAN-MP Wiki Calculations}\\
 \hline
 $\Delta x$ & \multicolumn{4}{|c|}{1 cm}\\
 $\Delta t$ & \multicolumn{4}{|c|}{$1.501 \times 10^{-11} s$}\\
 $n_{e^\pm}$ Species Num. Density & \multicolumn{4}{|c|}{$1 \times 10^{15} s$}\\
 $n_0$ Total Num. Density & \multicolumn{4}{|c|}{$2 \times 10^{15} s$}\\
 $w_0$ $\frac{Real Particles}{Single Macroparticle}$ & $3.125 \times 10^{13}$ & $3.125 \times 10^{13}$ & $1.5625 \times 10^{13}$ & $3.125 \times 10^{13}$\\
 Formula $\omega_{p,s}$ & \multicolumn{4}{|c|}{$1783255450012.7007 s^{-1}$}\\
 Formula $d_s$ &  \multicolumn{4}{|c|}{$2.5234746934141993e-13 cm   $}\\
 Fiducial $\omega_{e^-}^0$ & 0.0140625 & 0.0140625 & 0.0140625 & 0.1125 \\
 Fiducial $d_{e^-}^0$ & 32 & 32 & 32 & 4\\
 Actual $\omega_{p,s}$ & 0.0099 & 0.0099 & 0.0099 & 0.0795\\
 Actual $d_s$ & 45.2548 & 45.2548 & 45.2548 & 5.6568 \\
 \hline
\end{tabular}
\end{center}
\end{table}

\clearpage
\newpage

\section{\label{sec:level1}Characteristic Epochs}
\begin{table}[h]
\begin{center}
\caption{Characteristic epoch numbers and the corresponding behavior for the four simulation data sets.}
\begin{tabular}{ |p{4.5cm}||p{3cm}|p{3cm}|p{3cm}|p{3cm}|  }
 \hline
 \multicolumn{5}{|c|}{Characteristic Epochs} \\
 \hline
Simulation Parameter & S1 - Fiducial & S2 - Low $\gamma$ & S3 - High $PPC$ & S4 - Low $d_{e^-}$ \\
\hline
\hline
Thermal Noise & 1 & 1 & 1 & 1\\
Filament Ignition & 3 & 5 & 3 & N/A\\
Saturation & 5 & 7 & 5 & N/A\\
Filament Merger & 8 & 11 & 6 & N/A\\
Weibel Dissipation & 11 & 12 & 10 & N/A\\
Thermal/Numerical Noise & 15 & 15 & 15 & 15\\
\hline
\end{tabular}
\end{center}
\end{table}

\twocolumngrid
\nocite{*}
\bibliography{main}

\end{document}